\documentclass[prx,aps,twocolumn,superscriptaddress,floatfix,10pt]{revtex4-2}
\usepackage{standalone}
\usepackage{graphicx,color}

\input{scrload}
\usepackage{amsmath,bm}
\usepackage{amssymb}
\usepackage{nicefrac}
\usepackage{hyperref}

\usepackage{tikz}

\newcommand{\br}{{\bm r}}
\newcommand{\bu}{{\bm u}}
\newcommand{\bv}{{\bm v}}
\newcommand{\bs}{{\bm s}}
\newcommand{\bj}{{\bm j}}
\newcommand{\bq}{{\bm q}}
\newcommand{\bC}{{\bm C}}
\newcommand{\bV}{{\bm V}}
\newcommand{\ud}{{\mathrm d}}
\newcommand{\ue}{{\mathrm e}}

\newcommand{\bla}{{\big\langle}}
\newcommand{\bra}{{\big\rangle}}

\newcommand{\kb}{{k_\textsc{B}}}

\begin{document}

\title{\bf Temperature-driven flows in nanochannels: Theory and Simulations}

\author{Pietro Anzini}
\affiliation{Dipartimento di Scienza e Alta Tecnologia, Universit\`a degli Studi dell’Insubria, Via Valleggio 11, 22100 Como, Italy}
\affiliation{To.Sca.Lab, Universit\`a degli Studi dell’Insubria, Via Valleggio 11, 22100 Como, Italy}

\author{Zeno Filiberti}
\affiliation{Dipartimento di Scienza e Alta Tecnologia, Universit\`a degli Studi dell’Insubria, Via Valleggio 11, 22100 Como, Italy}

\author{Alberto Parola}
\email[Corresponding author: ]{alberto.parola@uninsubria.it}
\affiliation{Dipartimento di Scienza e Alta Tecnologia, Universit\`a degli Studi dell’Insubria, Via Valleggio 11, 22100 Como, Italy}
\affiliation{To.Sca.Lab, Universit\`a degli Studi dell’Insubria, Via Valleggio 11, 22100 Como, Italy}

\begin{abstract}
The motion of a fluid induced by thermal gradients in the absence of driving forces is known as thermo-osmosis. 
The physical explanation of this phenomenon stems from the emergence of gradients in the tangential pressure due to the presence of a confining surface. 
The microscopic origin of the effect was recently elucidated in the framework of linear response theory. Here, by use of conservation laws, we provide an explicit solution of the equations governing the fluid flow at stationarity in slab geometry, expressing the thermo-osmotic coefficient as the integrated mass current-heat current correlation function (which vanishes in the bulk). 
A very simple expression for the pressure gradient in terms of equilibrium properties is also derived. 
To test the theoretical predictions in a controlled setting, we performed extensive nonequilibrium molecular dynamics simulations in two dimensions.  Few simple models of wall-particle interactions are examined and the resulting pressure drop and velocity profile are compared with the theoretical predictions both in the liquid and in the gas regime. 
\end{abstract}
\maketitle

\section{Introduction}
Macroscopic fluid motion is always driven by pressure gradients or by external forces
and is accurately described by continuum approaches, like the Navier-Stokes equation~\cite{landau_fluid}. 
However, in sub-micron systems interfacial effects become relevant: Additional transport mechanisms are possible~\cite{surface_forces_1987} and the validity of the Navier-Stokes equations becomes questionable~\cite{bocquet_nanofluidics_2010}.

A remarkable example is provided by thermo-osmosis, a surface-induced phenomenon, where the external ``field'' driving the fluid flow is a temperature gradient.
The confinement of the fluid is essential for the onset of the flow.
Indeed, the steady state of a {\it bulk} fluid in a thermal gradient is characterized by the onset of local equilibrium:
All the macroscopic variables can be locally defined and the usual thermodynamic relations maintain their validity at each point. The temperature gradient thus leads to a spatial dependence of the chemical potential and of the particle density, keeping the bulk pressure uniform and preserving mechanical equilibrium. Thus, the fluid remains at rest while a stationary heat flux appears~\cite{landau_fluid,de_ma_1984}.
Only when a confining surface not orthogonal to the thermal gradient, say a wall, is present, the fluid starts moving, giving rise to thermo-osmosis.

Thermo-osmotic flows are expected to be minute on a macroscopic scale because the bulk fluid is set into motion due to an interfacial effect. Thus, the investigation of thermo-osmosis may appear to be justified mainly because it represents one of the simplest nonequilibrium systems we can envisage, serving as a test bed for any out-of-equilibrium theory of transport processes. 
However, the mass transport arising from any surface phenomenon becomes relevant every time one of the system's dimensions becomes comparable to the correlation length or to the mean free path of the fluid~\cite{osmosis_review_MB}. This situation occurs in several systems of technological and biological interest. Moreover, industrial miniaturization has made available a large variety of nanochannels, paving the way for the development of nanofluidics~\cite{squires_nanofluidics_2005} and many natural systems, such as membranes and gels, are characterized by porous networks with submicrometer diameters, where thermo-osmosis may play an important role~\cite{shukla_review_1984,barragankjelstrup_2016review}. In particular, in the field of low carbon energy conversion, researchers are investigating nanoporous membranes to transform waste heat into mechanical energy~\cite{liu_20,kim_mench_fuel}, and it is expected that the phenomenon will play an important role in the behaviour of clay-rich materials~\cite{1969_to_clay_dirksen, tremosa_2010_theo_insight,tremosa_2010_exp,to_clay_nuclear_2012}.
Furthermore, interest in thermo-osmosis is increasing with the possibility of applications as a mechanism for governing the motion of particles at the nanoscale~\cite{cichos_nanolett15,cichos_22,svetlana_2022,ruzzi_23}.

Thermo-osmosis in liquids was discovered more than a century ago~\cite{lippmann_1907,aubert_1912,bregulla_2016} and has been interpreted by Derjaguin on the basis of nonequilibrium thermodynamics, identifying the driving force as the local enthalpy excess induced by the confining surface~\cite{derjaguin_sidorenkov_1941thermoosmosis}, while, in rarefied gases, kinetic theories have been used since the pioneering works by Maxwell and Reynolds~\cite{feddersen_1972,reynolds_1879,maxwell_1879,kennard_1938kinetic,maxwell_reynolds_radiometer,piazza_2004}. 

However, the microscopic origin of the effect is still under debate, and the interpretation of nonequilibrium numerical simulations is carried out in the framework of phenomenological approaches, where additional free parameters must be introduced to provide a faithful description of the data~\cite{jivkov_21,jivkov_23,joly22,joly_23,joly_24,bjorn_2022,prezhdo_20,prezhdo_21,zambrano_23,TOslip_around_microparticle_2023}.

Recently, interest in thermo-osmosis has been revived by molecular dynamics simulations aimed at a microscopic understanding of this effect~\cite{ganti17,ganti18,proesmans_19,joly17,joly18}.
Following these investigations, a first principle treatment of thermal forces by use of linear response theory was developed~\cite{prl_2019,pre_2022}.
Explicit equations for the thermo-osmotic mass current were obtained in slab geometry, where the fluid is confined by two identical parallel walls, and a thermal gradient is imposed parallel to the surfaces. 
In the models studied in this work the walls are {\it passive}, that is they just act as an external force on the fluid molecules, orthogonal to the surface, thereby conserving the particle momentum parallel to the surface in a scattering process.

In this paper, using the continuity equations, we provide a general analytical solution of the equations developed in Refs.~\cite{prl_2019,pre_2022}, expressing the mass flow and the pressure drop induced by a thermal gradient in terms of the mass current-heat current correlation function without phenomenological fitting parameters. Our approach is valid at all density regimes, from gases to liquids. 
Taking advantage of this solution, we confirm the interpretation of thermo-osmosis as an effect of the variation of the local enthalpy density induced by the presence of the wall, making contact with previous phenomenological approaches~\cite{surface_forces_1987}, while providing the {\it unique} expression of the excess enthalpy, which, as stressed in Refs.~\cite{ganti17,henderson}, is not a well-defined quantity.

In order to test the theoretical predictions in a controlled environment, we present the results of extensive molecular dynamics simulations for a simple model of a fluid with one component and a few wall-particle interactions in two dimensions. 
Finally, we briefly consider the case of confining surfaces which violate the momentum conservation during the wall-molecule scattering, where the same microscopic theory allows to derive an analytical form of the velocity profile at low density.
Numerical and analytical results are compared, validating the theoretical interpretation in different regimes: From rarefied gases to dense liquids. 

The paper is organized as follows. 
The microscopic equations governing the fluid dynamics~\cite{prl_2019,pre_2022} are summarized in Section~\ref{sec:lrt}, where the explicit solution of the equations is also derived. 
In Section~\ref{sec:wp-int} the general solution previously found is specialized to different wall-particle interactions, while in Section~\ref{md}  we discuss the details of the equilibrium (MD) and nonequilibrium (NEMD) molecular dynamics simulations. The results are presented in Section~\ref{sec:results}, where particular attention has been devoted to the comparison between the outcomes of the theory and of the molecular dynamic simulations at several densities and temperatures, and for different wall-fluid potentials. 
Section~\ref{conclusions} contains some final remark. 

\section{Theory of thermo-osmosis}
\label{sec:lrt}
Thermo-osmotic effects have been known for more than a century and were
investigated in rarefied gases from the studies by Maxwell~\cite{maxwell_1879} and Reynolds~\cite{reynolds_1879}.
The microscopic understanding of this out-of-equilibrium phenomenon in gases,
first formulated by Maxwell~\cite{maxwell_1879} within kinetic theories, is quite subtle: The flow originates from the tangential stress on the gas layer near the confining surface due to the applied
temperature gradient. The effect strongly depends on the properties of the scattering event of a gas molecule on the wall, and notably on the momentum exchanged between the particle and the surface during the collision.
Remarkably, the flow disappears for an ideal gas confined by purely reflective hard walls: 
As originally noted in these studies, thermo-osmosis in rarefied gases needs some non-trivial wall structure. Elementary kinetic theory leads to an explicit expression for the slip velocity far from the interfaces, showing that the thermo-osmotic motion sets in along the direction of the temperature gradient, from the cold to the hot side~\cite{kennard_1938kinetic,sone_2000,sone_exp}.

Conversely, in the liquid regime, the phenomenological descriptions usually adopted rely on macroscopic approaches, such as nonequilibrium thermodynamics and Navier-Stokes equations~\cite{derjaguin_sidorenkov_1941thermoosmosis,surface_forces_1987,hutchisonnixondenbigh_1948,denbigh1952}.
However, the use of continuum theories can be justified when the relevant physical quantities
vary on a length scale much larger than the typical range of the interaction~\cite{galliero_2012,ganti17,ganti18}:
In the presence of a confining surface this condition is no longer satisfied, because, near the interface,
the fluid properties eventually driving the phenomenon may display strong, but short-ranged, modulations.

Only recently a first-principle and unitary theory of thermo-osmosis has been proposed~\cite{prl_2019}. This approach, valid both for gases and for liquids, is based on a generalization of the linear response theory formalism developed by Mori~\cite{mori58,balescu,zubarev1974nonequilibrium} to the case of an inhomogeneous fluid close to a surface.
This theory sheds light on the microscopic mechanisms driving thermo-osmosis and quantitatively relates the fluid flow to the properties of the fluid-wall interface via suitable dynamical correlation functions at equilibrium.

In slab geometry, namely when the fluid is confined between two infinite {\it passive} walls subject to a constant temperature gradient parallel to the walls, a chemical potential gradient, parallel to the temperature gradient, sets in together with a stationary mass flow.
The derivative of the velocity field with respect to the distance from the confining surfaces obeys a closed equation generalizing the macroscopic Stokes law to fluids confined in nanochannels.
In this ideal {\it open channel} configuration, where the confinement is due to a wall-particle force orthogonal to the walls, Galileo invariance along the direction of the gradient is preserved and then, the absolute fluid velocity is undetermined. In real systems, surface roughness imposes an additional no-slip boundary condition at the wall, forcing the vanishing of the mass current at contact. 

In the alternative {\it closed channel} geometry~\cite{pre_2022}, the number of particles is conserved and the total mass flow through the system vanishes, because of the additional hard wall boundary conditions imposed at the ``cold'' and ``hot'' sides. 
The thermo-osmotic flow near the two confining surfaces along the thermal gradient is compensated by a Poiseuille backflow at the center of the channel generated by a pressure imbalance between the hot and the cold side of the channel. This geometry is particularly suitable for numerical simulations and also allows for direct experimental investigations in membranes, aimed at the measure of the pressure difference at the two ends of the channel ~\cite{barragankjelstrup_2016review}. 

\subsection{Linear response theory}
In our system the fluid is confined in a slab centered in $z=0$ by an external, symmetrical ``wall potential" $V(z)=V(-z)$. 
The Hamiltonian defining the model is then
\begin{equation}
	\hat H = \sum_i \hat h_i = \sum_i \, \left [ \frac{p_i^2}{2m} + V(q_i^z) + \frac{1}{2} \sum_{j (\ne i)} v(q_{ij}) \right ],
	\label{ham}
\end{equation}
where $q_{ij} = |\bq_i - \bq_j|$ is the distance between particle $i$ and particle $j$. The slab extends to infinity in the $(x,y)$ plane: At equilibrium, translational invariance along this plane is
assumed and then the equilibrium properties of the fluid are uniform in these directions, while depend on the coordinate $z$. As a consequence, the $x$ and $y$ 
components of the total momentum of the particles are conserved by the microscopic dynamics. 
A temperature difference is set at the boundaries of the channel along the $x$ direction, leading to a uniform temperature gradient $\partial_x T$ throughout the channel. 
This problem was investigated by use of the linear response formalism~\cite{mori56, mori58, balescu, zubarev1974nonequilibrium}, which builds on the concept of local thermal equilibrium, defined by a Boltzmann weight
\begin{equation}
\ue^{-{\textstyle\int} \ud\br \,\beta(\br) \, \hat{\cal E}(\br)},
\notag
\end{equation}
where the energy density operator reads
\begin{equation}
	\hat{\cal E}(\br) = 
	\hat {\cal H}(\br) -\bu(\br)\cdot\hat \bj(\br) -\mu(\br)\hat \rho(\br).
        \notag
\end{equation}
The slowly varying external fields $\beta(\br)$, $\mu(\br)$ and $\bu(\br)$, related to the (inverse) temperature, chemical potential, and velocity profile, are parameters to be determined by the appropriate boundary conditions~\cite{pre_2022}.
$\hat {\cal H}(\br)$, $\hat \bj(\br)$ and $\hat \rho(\br)$ are the microscopic Hamiltonian, momentum and mass density operators respectively:
\begin{eqnarray}
        \hat {\cal H}(\br) &=& \sum_i \hat h_i \, \delta(\br - \bq_i); \notag \\
	\hat j^\alpha(\br) &=& \sum_i p^\alpha_i \,\delta(\br - \bq_i);  \label{massc}\\
        \hat \rho(\br) &=& m\, \sum_i \delta(\br-\bq_i). \notag
\end{eqnarray}
The steady state condition requires that the divergence of the mass flux, energy flux and momentum flux vanishes. These conditions translate in a set of integro-differential equations detailed in Ref.~\cite{prl_2019}.

In such a simple geometry, a solution has been found assuming that the only non-vanishing external parameters are $\partial_x\beta$ and $\partial_x(\beta\mu)$, independent of $z$ and $u^x(z)$. 
These equations were later extended to the case of a ``closed channel" geometry in~\cite{pre_2022}. Here Galileo invariance in the $x$ direction is broken by the boundary conditions which set a preferential reference frame and, at steady state, the fluid velocity sufficiently far from the hot and cold ends is fully determined by imposing the vanishing of the integrated mass flux, 
\begin{equation}
\int \ud z \, \langle j^x(z)\rangle = 0 \, ,
\label{eq:intj_closed}
\end{equation}
where the steady state average $\langle \, \cdot \, \rangle$ has been introduced in Ref.~\cite{prl_2019}.

Linear response theory provides the definition of the microscopic momentum and energy current operators $\hat J_j^{\alpha\nu}(\br)$ and $\hat J_H^\alpha(\br)$, which read~\cite{balescu}: 
\begin{eqnarray}
	\label{momc}
\hat J_j^{\alpha\nu}(\br) &=& \sum_i \left [ \frac{p_i^\alpha p_i^\nu}{m} \,\delta(\br-\bq_i) + \Gamma_i^{\alpha\nu}(\br)\right ] \\
	\label{energyc}
\hat J_H^{\alpha}(\br) &=& \sum_i \frac{p_i^\nu}{m} \,\left  [ \hat h_i\,\delta(\br-\bq_i) \delta^{\alpha\nu} +
\Gamma_i^{\alpha\nu}(\br)\right ], 
\end{eqnarray}
where the non-local contribution is given by
\begin{equation}
\Gamma_i^{\alpha\nu}(\br) = 
\frac{1}{2} \sum_{j(\ne i)} \frac{\partial v(q_{ij})}{\partial q_i^\alpha}\int_{C_{ij}}{\ud s^\nu} \,\delta(\br-\bs) 
\label{prob}
\end{equation}
and depends on the (arbitrary) choice of the path $C_{ij}$ connecting the position $\bq_i$ of particle $i$ to the position $\bq_j$ of particle $j$~\cite{henderson,rowlinson_1993}. 
The definition of heat flow requires the subtraction of the convective contribution from the microscopic 
expression of the energy flux~(\ref{energyc}). According to Ref.~\cite{prl_2019} we set:
\begin{equation}
	\hat J^x_Q(\br) = \hat J^x_H(\br) - \gamma\,\hat j^x(\br),
	\label{heatc1}
\end{equation}
where we introduced the ratio 
\begin{equation}
	\gamma = \frac{\partial_x [\beta\mu]}{\partial_x\beta}.
 \label{gammadef}
\end{equation}
This quantity is directly related to the uniform pressure gradient which develops in the $x$ direction
sufficiently far from the walls, where the fluid can be considered homogeneous and simple local equilibrium holds~\cite{pre_2022}:
\begin{eqnarray}
\partial_x p_b &=& \big[ \partial_\beta p_b\big\rvert_{\beta\mu} + \gamma\,\partial_{\beta\mu} p_b\big\rvert_{\beta} \big] \,\partial_x\beta \nonumber \\
        &=& \rho_b\,\kb T\,\big [ \gamma-h_m\big] \,\partial_x\beta,
\label{drop}
\end{eqnarray}
where $\rho_b$, $p_b$ and $h_m$ are the bulk mass density, bulk pressure and bulk enthalpy per unit mass, respectively evaluated at the average value of the (inverse) temperature and the chemical potential $\beta$ and $\mu$. 
The second line follows from standard thermodynamic identities.

In steady state conditions, the mass flux, in both open and closed channels, can be formally written as
the sum of three contributions~\cite{pre_2022}: 
\begin{eqnarray}
	\bla \hat j^x(z)\bra& =&\rho_0(z)u^x(z) - \nonumber \\
	&& \beta \, \int_0^\infty \ud t\int \ud\br^\prime \,
	\bla\hat j^x(\br,t)\hat J_j^{xz}(\br^\prime)\bra_0 \,\partial_{z^\prime}u^x(z^\prime) + \nonumber \\
	&& \int_0^\infty \ud t\int \ud\br^\prime \, 
	\bla\hat j^x(\br,t)\hat J_Q^x(\br^\prime)\bra_0 \,\partial_x\beta 
\label{masscurrent}
\end{eqnarray}
in terms of the thermal gradient $\partial_x\beta$ and of the velocity field $u^x(z)$. Here $\rho_0(z)$ is the equilibrium mass density and the symbol $\langle \, \cdot \, \rangle_0$ indicates averages evaluated at full thermodynamic equilibrium. 

While the thermal gradient is uniquely set by the temperature difference at the boundary and by the length $L_x$ of the channel, in order to fix the chemical potential gradient $\partial_x(\beta\mu)$ and the velocity field further equations are needed~\cite{prl_2019}.
Physically, they stem from the condition that the $x$ component of the average momentum density in the stationary state does not change in time: 
\begin{equation}
\label{gold}
\int \ud z^\prime \,\partial_z{\cal K}(z,z^\prime)\, \partial_{z^\prime}u^x(z^\prime) = \partial_z\big[{\cal S}_s(z)+
	{\cal S}_d(z)\big] \partial_x\beta \, .
\end{equation}
The kernel ${\cal K}(z,z^\prime)$ has the physical meaning of local viscosity
\begin{equation}
{\cal K}(z,z^\prime) = \beta \int \ud\br^\prime_\perp  \int_0^\infty \ud t^\prime  \bla\hat J_j^{xz}(\br,t^\prime)\hat J_j^{xz}(\br^\prime)\bra_0  \,
\label{cappa}
\end{equation}
while ${\cal S}_s(z)$ and ${\cal S}_d(z)$ are the static and dynamic source terms
\begin{eqnarray}
\partial_z {\cal S}_s(z) &=&
	-\int \ud\br^\prime \, \bla\hat j^{x}(\br)\,\hat J_Q^{x} (\br^\prime)\bra_0 \nonumber \\
	&=& - \kb T\,\big[ h_0^v(z) - \gamma\rho_0(z)\big]  \, ; \label{ss} \\
\partial_z {\cal S}_d(z) &=&\partial_z
\int_0^\infty \ud t \int \ud\br^\prime  \bla \hat J_j^{xz}(\br,t)\, \hat J^x_Q(\br^\prime) \bra_0 \, ,
\label{sd}
\end{eqnarray}
where $h_0^v(z)$ is the transverse component of the {\it virial} enthalpy density~(\ref{eq:virial_enthalpy}) in thermal equilibrium~\cite{pre_2022}. 
Equation~(\ref{gold}) closely resembles a linearized ``Navier-Stokes-like" equation where the viscosity coefficient $\eta$ is replaced by a non-local viscosity kernel ${\cal K}(z,z^\prime)$~(\ref{cappa}).

\subsection{A general solution for the mass flow}
\label{massflow}
The set of equations describing the thermo-osmotic flow~(\ref{masscurrent}-\ref{sd}) 
is clearly quite complex and its solution requires the knowledge of several dynamic correlation functions at 
equilibrium. Remarkably, conservation laws allow to considerably simplify the evaluation of 
the mass flow~(\ref{masscurrent}). In addition, the chemical potential gradient and the parameter $\gamma$ (\ref{gammadef}) entering the expression of the heat current can be fixed by imposing the (physical) constraint of finite mass flux providing an explicit solution to the problem. 

\subsubsection*{Continuity equations}
The key ingredient is the microscopic continuity equation obeyed by the momentum density operator by virtue of Hamilton equations: 
\begin{equation}
        \frac{\ud}{\ud t} \hat j^\alpha(\br) + \partial_\nu \hat J_j^{\alpha\nu}(\br) =  -\frac{\hat \rho(\br)}{m} \,\partial_\alpha V(z).
        \label{cont}
\end{equation}
It is convenient to define two spatially-integrated, time-dependent quantities: 
\begin{eqnarray}
	\label{f}
	F(z,t) &=& \int \ud\br^\prime \, \bla \hat j^x(\br,t)\hat J_Q^{x}(\br^\prime)\bra_0; \\
	\label{g}
	G(z,t) &=& \int \ud\br^\prime \, \bla \hat J_j^{xz}(\br,t)\hat J_Q^x(\br^\prime)\bra_0 .
\end{eqnarray}
The continuity equation~(\ref{cont}) sets a close relationship between these two functions:
\begin{equation}
        \frac{\partial F(z,t)}{\partial t} = - \frac{\partial G(z,t)}{\partial z}.
        \label{fg}
\end{equation}
As a consequence, a globally conserved quantity can be defined:
\begin{equation}
	\frac{\ud}{\ud t} \int \ud z\,F(z,t) = \int \ud z\,\partial_z G(z,t) = 0 
	\label{f00}
\end{equation}
because $G(z,t)$ must vanish at $z\to\pm\infty$ due to the presence of a confining potential.  
Therefore, the spatial integral of $F(z,t)$ is time-independent. Setting $t=0$ we find 
\begin{eqnarray}
	\int \ud z\,F(z,t) &=& \int \ud z\,F(z,0) \nonumber \\
	&=& \int \ud z \int \ud \br^\prime \, \bla \hat j^x(\br) \hat J^x_Q(\br^\prime) \bra_0 \nonumber \\
        &=& \kb T \int \ud z \, \big[ h_0^v(z) - \gamma\rho_0(z)\big] \, ,
	\label{f0}
\end{eqnarray}
where the last line was demonstrated in Ref.~\cite{pre_2022} and the definition of the virial enthalpy density $h_0^v(z)$ is given in Eq.~(\ref{eq:virial_enthalpy}).

Let us now consider the term 
\begin{equation}
	L(z) = \int_0^\infty \ud t \int \ud  z^\prime \int \ud  \,\br_\perp^\prime\bla 
	\hat j^x(\br,t) \hat J_j^{xz}(\br^\prime)\bra_0 \,\partial_{z^\prime} u^x(z^\prime)
	\nonumber 
\end{equation}
appearing in Eq.~(\ref{masscurrent}). Integrating by parts in $z^\prime$ and using the translational invariance in the $(x,y)$ plane, $L(z)$ becomes
\begin{eqnarray}
	L(z) &=& -\int_0^\infty \!\! \ud t \int \ud z^\prime\, u^x(z^\prime) \, \partial_{\alpha^\prime}\!\!  \int \ud \br_\perp
	\bla \hat j^x(\br,t)\hat J_j^{x\alpha}(\br^\prime)\bra_0 \nonumber \\
	 &=&\int_0^\infty \!\! \ud t \int \ud z^\prime\, u^x(z^\prime)\, \partial_{\alpha^\prime }\!\! \int \ud \br_\perp
        \bla \hat j^x(\br) \hat J_j^{x\alpha}(\br^\prime,t)\bra_0\,,
        \nonumber
\end{eqnarray}
where the second line follows from the known, general properties of the dynamical correlation functions at equilibrium~\cite{simpleliquids_fourth}. 
By use of the continuity equation~(\ref{cont}) the spatial derivative is converted into a time derivative and the integral 
over time is easily carried out 
\begin{eqnarray}
	L(z) &=& \int \ud \br^\prime \bla \hat j^x(\br)\hat j^x(\br^\prime)\bra_0 u^x(z^\prime) - \nonumber \\
	&& \qquad \lim_{t\to\infty} \int \ud \br^\prime \bla \hat j^x(\br) \hat j^x(\br^\prime,t)\bra_0 u^x(z^\prime)\nonumber \\
	=\, &\beta^{-1}&\rho_0(z)u^x(z)-\lim_{t\to\infty} \int \ud \br^\prime \bla \hat j^x(\br) \hat j^x(\br^\prime,t)\bra_0 u^x(z^\prime)\,.
	\nonumber 
\end{eqnarray}
Inserting this result into the expression for the mass current~(\ref{masscurrent}) we finally get
\begin{align}
	\bla \hat j^x(z)&\bra = \int_0^\infty  \ud t\int \ud\br^\prime \, 
        \bla\hat j^x(\br,t)\hat J_Q^x(\br^\prime)\bra_0 \,\partial_x\beta \nonumber \\
	& + \beta\lim_{t\to\infty} \int \ud \br^\prime \bla \hat j^x(\br)\hat j^x(\br^\prime,t)\bra_0 \,u^x(z^\prime) \, .
\label{masscurrent3}
\end{align}

\subsubsection*{The constraint of finite integrated mass flux}
We can now evaluate the {\it spatially-integrated} mass flux, that is the total mass flowing through a section of the channel per unit time. 
By integrating Eq.~(\ref{masscurrent3}) in $z$, 
the first contribution coincides with the time integral of the conserved (i.e. time-independent) quantity~(\ref{f0}), 
which must then vanish in order to give a finite result:  A finite integrated mass flux requires that 
$\int \ud z \,F(z,t)=0$, providing, via Eq.~(\ref{f0}), the value of the parameter $\gamma$: 
\begin{equation}
        \gamma = \frac{\int \ud z \,h^v_0(z)}{\int \ud z \,\rho_0(z)} \, .
        \label{gamma}
\end{equation}
In the bulk limit, both the enthalpy density and the mass density are uniform and then $\gamma$ equals the enthalpy per unit mass $h_m$.
This limit is reached in a semi-infinite system confined by a single wall. 

\subsubsection*{Mass flow}
Finally, we turn our attention to the velocity field $u^x(z)$. 
First we note that, by virtue of the previous definitions, the derivative 
of the dynamical source term~(\ref{sd}) can be written as
\begin{eqnarray}
	\partial_z {\cal S}_d(z) &=& \int_0^\infty \ud t \,\partial_z G(z,t) = F(z,0) - 
	\lim_{t\to\infty} F(z,t) \nonumber \\
	&=& -\partial_z {\cal S}_s(z)-\lim_{t\to\infty} F(z,t) \nonumber \\
	&=& -\partial_z {\cal S}_s(z),
	\label{drama}
\end{eqnarray}
where we used the continuity equation~(\ref{fg}) in the first line and Eq.~(\ref{ss}) in the second. The final result (\ref{drama}) holds because the mass current-heat current integrated dynamical correlation $F(z,t)$ tends to zero at large times, to guarantee the convergence of the time integral in Eq.~(\ref{masscurrent3}). 
Therefore the source term in the ``Stokes" equation~(\ref{gold}), given by the sum of the static and the dynamical contribution, 
identically vanishes, meaning that the velocity field must be a constant
\begin{equation}
u^x(z)=u^x_0 \, .
\notag
\end{equation}
This does not imply that the velocity profile is flat: As shown in Eq. (\ref{masscurrent3}), the velocity field $u^x(z)$ does not fully determine the mass flow.

Substituting this result into the expression for the mass current~(\ref{masscurrent3}), we get 
\begin{align}
	\bla \hat j^x(z)\bra = u^x_0 & \,\beta\lim_{t\to\infty} \int \ud \br^\prime \bla \hat j^x(\br) \hat j^x(\br^\prime,t)\bra_0 + 
	\nonumber \\ 
	& \int_0^\infty \ud t\int \ud\br^\prime \, \bla\hat j^x(\br,t)\hat J_Q^x(\br^\prime)\bra_0 \,\partial_x\beta.
\notag
\end{align}
The first term is easily evaluated thanks to the conservation of the $x$ component of the total momentum: 
\begin{equation}
    \beta \! \int \! \ud \br^\prime \bla \hat j^x(\br)\hat j^x(\br^\prime,t)\bra_0 = \beta
    \!\! \int \! \ud \br^\prime \bla \hat j^x(\br)\hat j^x(\br^\prime)\bra_0 =
    \rho_0(z),
	\nonumber 
\end{equation}
that holds at all times, and then the general expression for the mass current reads:
\begin{align}
	\bla \hat j^x(z)\bra = \,&\rho_0(z) \, u^x_0 \,\,+ \notag \\
         &\int_0^\infty \ud t\int \ud\br^\prime \bla\hat j^x(\br,t)\hat J_Q^x(\br^\prime)\bra_0\partial_x\beta.
\label{jx}
\end{align}
In agreement with the linear response approach, the thermo-osmotic flow is proportional to the temperature gradient $\partial_x\beta$. 

\subsubsection*{Pressure gradient}
As shown in Appendix~\ref{app-virial}, the spatial integral of the transverse virial enthalpy density appearing in Eq.~(\ref{gamma}) coincides with the thermodynamic enthalpy per unit area of the confined fluid.
Therefore, the general expression for the pressure gradient at the center of the channel~(\ref{drop}) can be written as
\begin{equation}
	\partial_x p_b = -\rho_b\,\Delta h_m\frac{\partial_x T}{T},
	\label{drop2}
\end{equation}
where $\Delta h_m$ is the difference between the average enthalpy per unit mass in the channel and
its bulk value. This expression closely resembles the result obtained in Refs.~\cite{derjaguin_sidorenkov_1941thermoosmosis,surface_forces_1987} in narrow pores by use of nonequilibrium thermodynamics. The microscopic approach presented here provides the precise definition of macroscopic quantities like the ``mean, or effective, value of the excess enthalpy"~\cite{surface_forces_1987} within the realm of equilibrium statistical physics. 

\subsection{Velocity profile}
At this point the problem of thermo-osmosis is not completely solved: The mass current~(\ref{jx}) is still undetermined because the velocity shift $u^x_0$ has not been fixed yet.
In an {\it open} configuration the absolute fluid velocity can not be fixed uniquely, due to Galileo invariance of the system that must be preserved in our model. On the other hand, in a {\it closed} channel this parameter is determined by imposing the vanishing of the integrated mass flux~(\ref{eq:intj_closed}). 

As previously shown, the $z$-integral of the second contribution in Eq.~(\ref{jx}) identically vanishes due to the choice of $\gamma$~(\ref{gamma}) and then
the integrated mass flux just becomes 
\begin{equation}
\int\ud z \,\bla \hat j^x(z)\bra = u^x_0 \,\kb T\int\ud z \,\rho_0(z) \, .
\end{equation}
In a closed channel Eq.~(\ref{eq:intj_closed})
implies $u^x_0=0$, leading to the final expression of the mass flux: 
\begin{equation}
	\bla \hat j^x(z)\bra = 
	\int_0^\infty \ud t\int \ud\br^\prime \, \bla\hat j^x(\br,t)\hat J_Q^x(\br^\prime)\bra_0 \,\partial_x\beta \, .
	\label{masscurfin}
\end{equation}
This equality shows that the mass current in a closed system is ruled by the relevant Onsager coefficient (thermo-osmotic coefficient) and proportional to the temperature gradient.

\subsubsection*{The thermo-osmotic coefficient vanishes in bulk}
Thermo-osmosis is an interface-driven effect occurring only in confined systems: Neither fluid motion, nor a pressure gradient can be observed in a homogeneous system. In bulk, the excess enthalpy vanishes by definition and therefore the pressure drop in Eq.~(\ref{drop2}) disappears. Analogously, also the thermo-osmotic coefficient appearing in Eq.~(\ref{masscurfin}) must vanish when translational invariance holds. This result, usually quoted in the literature without any proof at the microscopic level~\cite{balescu}, can easily be recovered within our formalism. 

By dropping the integration in $z^\prime$ in the definitions (\ref{f},\ref{g})
and following the same argument as above, we show that the quantity $\int \ud \br \, \bla \hat{j}^x(\br,t)\,\hat{J}^x_Q(\br^\prime) \bra_0$
is time-independent and in bulk vanishes at all times, due to the chosen value of $\gamma$. This implies that also its time integral is
identically zero, proving that in bulk the mass current-heat current transport coefficient vanishes~\cite{balescu}:
\begin{equation}
	\int_0^{\infty} \ud t \int \ud \br \, \bla \hat{j}^\alpha(\br,t)\,\hat{J}^\beta_Q(\br^\prime) \bra_0 = 0 \, .
	\label{eq:plati}
\end{equation}

\section{The wall-particle interaction}
\label{sec:wp-int}
Having shown that the presence of a confining surface is essential for triggering the thermo-osmotic flow, we now examine two different families of walls where some approximate result can be obtained. The first Subsection is devoted to the study of passive walls, both in the rarefied and in the dense fluid case, whereas the second one to the so-called diffusive wall proposed by Maxwell.
In the gas regime we will assume that the fluid is made up of hard particles of diameter $\sigma$.

\subsection{Passive walls}
First we discuss the approximate evaluation of the mass flux~(\ref{masscurfin}) in the case of a confining surface preserving the conservation of the component of the particle momentum parallel to the wall during the scattering process. 
The gas and liquid regimes are addressed separately in the two Subsections below. 

\subsubsection*{The gas regime}
\label{gasreg}
In the low density limit, where the mean free path 
$\lambda = 1/\big(\sqrt{2}\,n_0\,\pi\sigma^2\big)$ is comparable to the size of the slab, the 
Knudsen number $\mathrm{Kn} =\lambda / L_z \gtrsim 0.1$ and standard kinetic theory can be used to evaluate the dynamical correlation function. 
Disregarding the interaction term in the definition of the heat flux, the relevant dynamic correlation function appearing in~(\ref{masscurfin}) reads:
\begin{eqnarray}
	&&\int \ud\br^\prime \, \bla\hat j^x(\br,t)\hat J_Q^x(\br^\prime)\bra_0 = \nonumber \\
	&&\quad\quad \frac{1}{Am} \sum_{i,j} \left\langle p^x_i(t) \, \delta\big(z-q^z_i(t)\big) p^x_j \Big[ \hat h_j - \gamma m\Big] \right\rangle_0,
	\label{kin1}
\end{eqnarray}
where $A$ is the total area in the $(x,y)$ plane, we used translational invariance at equilibrium
and the single particle Hamiltonian is just $\hat h_j = \nicefrac{p^2_j}{2m} + V(q^z_j)$. Introducing the 
relaxation time approximation~\cite{reif2009fundamentals}, we assume that particles are mutually uncorrelated and the single particle dynamics 
is ballistic up to a collision event, while after collision the memory is lost. 
The probability of interparticle scattering in the time interval $\ud t$ is just $\ud t / \tau$, where $\tau$ is the collision time.

Under these hypotheses, the dynamical correlation~(\ref{kin1}) becomes
\begin{equation}
	\frac{N(t)}{A m} 
	\left\langle p^x(t)\delta(z-z(t)) p^x\left [ \hat h - \gamma m\right ] \right\rangle_0,
 \notag
\end{equation}
where $N(t) = N\,\ue^{-t/\tau}$ is the number of particles which did not experience collisions up to a time $t$. 
Now we recall that both the $x$ component of the momentum $p^x$ and the total energy $\hat h$ are conserved in the single-particle ballistic dynamics. Then we can evaluate all the terms in this expression at the same time $t$. In this way, the dynamical correlation reduces to a static one, which is easily evaluated: 
\begin{equation}
	\ue^{-t/\tau}n_0(z) \, \kb T\,\left [ \frac{5}{2} \kb T + V(z) - \gamma m\right ],
 \notag
\end{equation}
where $n_0(z)$ is the number density. 
Note that the low density limit of the local enthalpy is just $h_0^v(z) = [ \nicefrac{5}{2} \, \kb T + V(z)] \, n_0(z)$
and then the integral of this correlation over $z$ vanishes at all times with the adopted choice~(\ref{gamma}) of the parameter $\gamma$, 
in agreement with the general result~(\ref{f0}). Integrating over time we obtain the mass flux in the gas regime: 
\begin{equation}
	\langle j^x(z)\rangle = \tau\,\kb T\,\big[ h_0^v(z)- \gamma \rho_0(z)\big]\partial_x \beta \, .
	\label{low}
\end{equation}
The evaluation of the collision time $\tau$ according to elementary kinetic theory is reported in Appendix~\ref{app-collision}. 

Equation~(\ref{low}) predicts that
at low density the flux becomes of the ``plug" type far from the confining surfaces, where the wall-particle interaction vanishes. 
The velocity of the fluid becomes uniform in $z$ and the mass flows in the same direction of the temperature gradient for repulsive potentials and opposite for attractive interactions while
near the walls the velocity changes sign to respect the global requirement of mass conservation.
Note that in the limit of a purely hard-wall interaction, the fluid velocity vanishes identically at low density because of the exact compensation between the space-independent enthalpy per unit mass  
and the term $\gamma\rho$.

\subsubsection*{The liquid regime}
In a dense fluid, where $\mathrm{Kn} \lesssim 0.01$,
correlations play a crucial role, kinetic theory loses accuracy and 
the evaluation of the mass flux should be performed by use of the hydrodynamic approximation.
At equilibrium no thermal gradient is present and then we assume that the main contribution to the energy flux  
$\hat J_H^x(\br)$~(\ref{energyc}) is given by the convective term:
\begin{equation}
        \hat J^x_H(\br) \approx \frac{h_0^v(z)}{\rho_0(z)}\,\hat j^x(\br) \, .
    \notag
\end{equation}
Note that, according to our definition~(\ref{heatc1}) of heat flux $\hat J_Q^x(\br)$, only the 
spatial average of the convective energy current is subtracted: While in a homogeneous fluid this procedure reproduces
the standard definition of heat flux, in the presence of inhomogeneities a residual convective contribution remains. 
Accordingly, the integrated mass current-heat current dynamical correlation ruling thermo-osmosis will be evaluated as:
\begin{eqnarray}
	&&\int\ud\br^\prime\,\bla \hat j^x(\br,t)\, \hat J^x_Q(\br^\prime) \bra_0 \approx \nonumber \\
	&&\qquad \quad \int\ud\br^\prime\,
	\left [ \frac{h_0^v(z^\prime)}{\rho_0(z^\prime)} - \gamma \right ] \,\bla \hat j^x(\br,t)\, \hat j^x(\br^\prime) \bra_0.\quad
	\label{dress}
\end{eqnarray}
This approximation satisfies the requirement~(\ref{f0}), verifying the condition of vanishing integrated flux for the choice~(\ref{gamma}) of the parameter $\gamma$. 

The integrated transverse current correlation function, defined as
\begin{equation}
     C_t(z,z^\prime) = \int \ud\br^\prime_\perp \bla \hat j^x(\br,t)\,\hat j^x(\br^\prime)\bra_0,
\notag
\end{equation}
can be estimated by use of the hydrodynamical approximation~\cite{simpleliquids_fourth} usually adopted in bulk fluids and extended to confined fluids in slab geometry in Refs.~\cite{boba_prl,boba}.
A reliable approximation should preserve the identity $\int \ud z \,C_t(z,z^\prime) = \kb T\rho_0(z^\prime)$ which guarantees the vanishing of the integrated mass flux. Following Ref.~\cite{boba}, we first subtract an $O(1/L_z)$ contribution by setting

\begin{equation}
	C_t(z,z^\prime) = \kb T \, \frac{\rho_0(z)\rho_0(z^\prime)}{\int \!\ud \zeta \,\rho_0(\zeta)} + \bar C_t(z,z^\prime).
 \notag
\end{equation}
Note that the time-independent shift does not contribute to Eq.~(\ref{dress}) due to the choice~(\ref{gamma}) of the parameter $\gamma$. 
The correlation function $\bar C_t(z,z^\prime)$, which must vanish upon integration in $z$, 
is evaluated by use of the approximation derived in Ref.~\cite{boba}. Integrating in time, the result reads
\begin{equation}
	\int_0^\infty  \ud t \,\bar C_t(z,z^\prime) = \frac{\rho_b \rho_0(z^\prime)}{2\eta\beta}
	\left[ \frac{L_z}{6} + \frac{z^2 + {z^\prime}^2}{L_z} - |z-z^\prime|\right],
	\nonumber 
\end{equation}
where $\eta$ is the bulk shear viscosity.
When this expression is inserted into Eqs.~(\ref{masscurfin},\ref{dress}), it provides the mass flux in hydrodynamical approximation:
\begin{eqnarray}
	\bla \hat{j}^x(z)\bra &=& -\frac{\rho_b \partial_x T}{2\eta T}\, \int \ud z^\prime \,
        \big[ h_0^v(z^\prime) -\gamma\rho_0(z^\prime)\big]
	\times \nonumber \\
	&&\qquad\quad  \left [ \frac{L_z}{6} + \frac{z^2 + {z^\prime}^2}{L_z} - |z-z^\prime|\right ]. 
	\label{massflux}
\end{eqnarray}

Taking the second spatial derivative of the mass flux~(\ref{massflux}) we obtain an expression that closely resembles the Stokes equation: 
\begin{equation}
	\frac{\ud^2 \bla \hat{j}^x(z)\bra}{\ud z^2} = \frac{\rho_b\partial_x T}{\eta T} \big[ h_0^v(z)-\gamma\,\rho_0(z)\big].
	\label{nse}
\end{equation}
Far from the walls, the fluid is homogeneous and Eq.~(\ref{nse}) reduces to the known parabolic Poiseuille flow
\begin{equation}
	\frac{\ud^2 v^x(z)}{\ud z^2} = \frac{\rho_b\partial_x T}{\eta T} \big[ h_m-\gamma\big],
 \notag
\end{equation}
consistent with the general expression~(\ref{drop}) of the pressure gradient far from the walls.

In the case of a single confining surface at $z=0$ the parameter $\gamma$ equals $h_m$, and Eq.~(\ref{massflux}) reduces to:
\begin{equation}
	\bla \hat{j}^x(z) \bra \!= \! {j}_0^x \,
	-\,\frac{\rho_b\partial_x T}{\eta T}\!\!\int_0^\infty \!\!\!\!\!\ud z^\prime \,{\rm Min}(z,z^\prime)\big[ h_0^v(z^\prime)-h_m\,\rho_0(z^\prime)\big],
	\nonumber
\end{equation}
where we introduced the shift $j_0^x$ because in a fluid confined by a single surface the constraint of 
vanishing integrated mass flux does not apply. In this case, the shift is determined by the boundary condition 
at the wall: No-slip boundary conditions correspond to the choice $j^x_0=0$. 
Under this boundary condition the corresponding asymptotic velocity far from the wall reads, 
\begin{equation}
        v^x_\infty= 
        -\frac{\partial_x T}{\eta T}\int_0^\infty \ud z\, z\,\big[ h_0^v(z)-h_m\,\rho_0(z)\big],
	\notag
\end{equation}
providing the expression for the thermo-osmotic coefficient~\cite{ganti17}:
\begin{equation}
	\beta_{12}=-\frac{v^x_\infty}{\partial_x T/T}=
        \frac{1}{\eta}\int_0^\infty \ud z\, z\,\big[ h_0^v(z)-h_m\,\rho_0(z)\big].
        \label{beta12}
\end{equation}

\subsection{Diffusive wall}
 
Now we briefly consider the case of walls violating the conservation of the particle momentum parallel to the surface during the wall-molecule scattering. The Hamiltonian of the system is still defined by~(\ref{ham}), and therefore the continuity equations still hold within the fluid. Only the evaluation of the dynamical correlation in Eq.~(\ref{jx}) and the determination of
the constant $\gamma$ and of the velocity shift $u^x_0$ will be affected by the peculiarities of the wall-molecule scattering. 
As an illustrative example we consider a rarefied gas confined by the so called ``diffusive walls".
As suggested by Maxwell~\cite{maxwell_1879}, real surfaces are characterized by the presence of roughness,
which can be modeled by small asperities with different heights: These asperities entangle the impinging particles, which,
after some collisions in the cavities, return in the surrounding gas with a different momentum. This behavior can be
modeled by the so called diffusively reflecting wall. The diffusive wall confines in a half-space the fluid,
exactly like a flat hard wall: The fraction $1-f$ of the particles is reflected, while the fraction $f$ is absorbed and afterwards evaporated with a new momentum selected from a Maxwell-Boltzmann distribution at the local temperature of the wall. 

In order to investigate the  effects of diffusive surfaces on the thermo-osmotic slip within our formalism, 
we define the scattering process so that the normal component $p^z$ of the momentum of the impinging particle is specularly reflected after the collision, as is the case of a smooth hard wall, while the tangential components after the collision are isotropically distributed, preserving the conservation of the particle kinetic energy. 
In this way, the energy of the fluid is strictly conserved by the dynamics and only the momentum conservation is violated. 

\begin{figure}
    \input{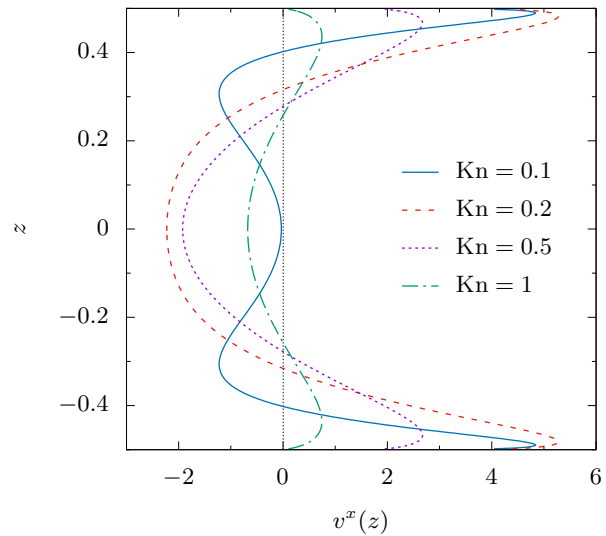}
    \caption{Thermo-osmotic velocity profile of a gas in a closed channel of width $L_z$ bounded by  diffusive walls. Velocities are evaluated via Eq.~(\ref{veldif}) and are expressed in units of $\tau \, \frac{\kb T}{m}\,\frac{\partial_x T}{T}$. The coordinate $z$ is expressed in units of $L_z$. Colors refer to different channel widths, corresponding to Knudsen numbers ranging from $0.1$ to $1$.}
    \label{fig-velchan}
\end{figure}

In rarefied gases, an estimate of the mass flux can be obtained by use of elementary kinetic theory in Eq.~(\ref{jx}).
Under the relaxation time approximation, detailed in Sect.~\ref{gasreg}, the calculation of the dynamical correlation function, sketched in Appendix~\ref{app-diffusive}, leads to an explicit expression for the fluid velocity in a channel of width $L_z$: 
\begin{eqnarray}
	\label{veldif}
	&& v^x(z) = u_0^x + 
        \frac{(\kb T)^2}{m\sqrt{\pi}} \,\partial_x\beta 
	\int_0^\infty \ud t\,\ue^{-t/\tau} \\
	&& \left \{ \left [ 2-\beta\gamma m
        +\zeta_+^2\right ] \,\zeta_+\ue^{-\zeta_+^2} + 
        \left [ 2-\beta\gamma m
	+\zeta_-^2\right ] \,\zeta_-\ue^{-\zeta_-^2} \right \}.
	\nonumber 
\end{eqnarray}
Here, $\zeta_\pm^2 = \beta m (L_z\pm 2z)^2 / \big(8t^2\big)$ and we determined the parameter $\gamma$ and the velocity shift $u_0^x$ entering this expression according to Eqs.~(\ref{eqgamma}) and (\ref{uzero}),  
derived in Appendix~\ref{app-diffusive}. A plot of the velocity profile in channels of different widths is shown in Fig.~\ref{fig-velchan}. 

\section{Molecular Dynamics simulations}
\label{md}
The model system for the study of thermo-osmosis is an infinite slit pore 
filled with a fluid, where a temperature gradient is imposed in the direction parallel to the boundaries.
The symmetry of the problem suggested to perform a nonequilibrium molecular dynamics simulation of a {\it two}-dimensional nanochannel (see Fig.~\ref{fig:sys}), which retains all the key physical
features at the basis of this effect and, at the same time, reduces the number of particles involved and then the computational cost, allowing for higher accuracy.
Following a pioneering investigation~\cite{bjorn1999} we performed 
nonequilibrium simulations of a fluid confined in a {\it closed} channel subject to a thermal gradient. This set up is particularly convenient because it does not require the use of periodic boundary conditions along the temperature gradient and leaves the system free to adjust the pressure gradient along the channel. 
The main outputs of a nonequilibrium simulation of thermo-osmosis are the velocity profile and the pressure gradient along the channel, an observable easily accessible in experiments ({\it e.g.} in membranes).

The theory presented in the preceding Sections provides explicit, parameter-free, expressions for the pressure gradient and the velocity profile that require the evaluation of the density and the enthalpy profile in the fluid, as well as the bulk viscosity. For this reason we also performed equilibrium molecular dynamics simulations of the same system to access these observables. 

Here we introduce all the details shared by both equilibrium and nonequilibrium simulations, like the particle interaction and the confining potentials.
In all the simulations we performed the fluid particles interact through a truncated and shifted 12/6 Lennard-Jones (LJ) potential
\begin{equation}
v(r) =
    \begin{cases}
            v_{\mathrm{LJ}}(r) - v_{\mathrm{LJ}}(r_c) & r \leq r_c\\
            0 & r > r_c
    \end{cases},
    \label{eq:ffpot}
\end{equation}
where the expression of the LJ potential reads
\begin{equation}
v_{\mathrm{LJ}}(r) =4\epsilon \left[ \left( \frac{\sigma}{r} \right)^{12} - \left( \frac{\sigma}{r} \right)^6 \right].
\label{eq:LJ126}
\end{equation}
The parameters $\epsilon$ and $\sigma$ are the depth of the potential well
and the particle diameter respectively, whereas the
cutoff radius $r_c$ is set to $4.5\sigma$. The dimensional constants $\sigma$ and $\epsilon$, together with the
particle mass $m$ allow to define the standard time unit $\tau = \sigma\sqrt{{m}/{\epsilon}}$.

The role of confinement has been numerically investigated by considering
different walls in the direction perpendicular to the flow. 
The simplest choice is to confine the fluid in the region $- {L_z}/{2} < z < {L_z}/{2}$ with a couple of reflective walls, described by the external potential
\begin{equation}
V_{hw}(z) = W_{hw}\left(z+\frac{L_z}{2}\right) + W_{hw}\left(\frac{L_z}{2}-z\right),
\notag
\end{equation}
where
\begin{equation}
        W_{hw}(d) =
    \begin{cases}
            +\infty     &d \le 0\\
            0          &d > 0
    \end{cases}.
    \label{eq:HW}
\end{equation}
To model a more physical confinement, where the effect of the wall extends also in the fluid region, we 
added to the bare hard wall potential two different tails with a finite, non-zero range, characterized by
an overall attractive and repulsive behavior.
The first one is a purely repulsive interaction
\begin{equation}
        W_{r}(d) =
    \begin{cases}
            k(d-d_r)^2   &0<d \leq d_r\\
            0               & d > d_r
    \end{cases},
    \notag
\end{equation}
where $k = 0.1 \epsilon/ \sigma^{2}$ and $d_r = 5\sigma$.
The resulting repulsive external potential $V_{r}(z)$ is therefore given by 
\begin{equation}
V_{r}(z) = V_{hw}(z)+W_{r}\left(z+\frac{L_z}{2}\right) + W_{r}\left(\frac{L_z}{2}-z\right).
\label{eq:Vr}
\end{equation}
The second one is defined by the (truncated) Lennard-Jones 9/3 form:
\begin{equation}
        W_{a}(d) =
    \begin{cases}
            \epsilon_{93} \left[ \frac{2}{15} \left( \frac{\sigma_{93}}{d} \right)^9 - \left( \frac{\sigma_{93}}{d} \right)^3 \right] & 0 < d \leq d_a\\
            0 & \qquad d > d_a
    \end{cases},
    \notag
\end{equation}
where $\epsilon_{93} = \epsilon$, $\sigma_{93} = \sigma$ and $d_a = 10\sigma$.
This potential is attractive everywhere except for $d < \left( {2}/{5} \right)^{{1}/{6}}\sigma$,
where it becomes strongly repulsive.
The contribution of this tail gives rise to the so called attractive potential given by
\begin{equation}
V_{a}(z) = V_{hw}(z)+W_{a}\left(z+\frac{L_z}{2}\right) + W_{a}\left(\frac{L_z}{2}-z\right).
\label{eq:Va}
\end{equation}

The last kind of confining potential we introduced in the preceding Section is the diffusive wall, denoted by $V_d$.
Our custom implementation in LAMMPS of this wall is slightly different than Maxwell’s original idea: The normal component of the impinging particle is specularly reflected after the collision, as in the case of a smooth hard wall, but the tangential component $p^x$ is taken from a Maxwell-Boltzmann distribution at the local temperature $T(x)$ of the wall, defined by
\begin{equation}
T(x)=T_C+\frac{T_H-T_C}{L}x \, ,
\notag
\end{equation}
where $L$ is the length of the system outside the reservoirs and $x\in [0,L]$ is the coordinate of the point of impact (see Fig.~\ref{fig:sys}). The diffusive confining potential, employed for the first time in a molecular dynamics simulation many years ago~\cite{suh_macelroy_86}, has been applied to a nonequilibrium molecular dynamics simulation in Ref.~\cite{bjorn1999}.

All the molecular dynamics simulations were performed through the LAMMPS~\cite{LAMMPS} package (\texttt{http://lammps.sandia.gov}), supplemented by few customized routines.

\subsection{Equilibrium molecular dynamics simulations}
The simulation box is a standard two-dimensional rectangular box, periodic in the $x$ direction, of width $L_z=30\sigma$ and length $L_x$ ranging from $60$ to $5000$, depending on the density (to increase the statistics we simulated bigger systems at the lower densities). In the $z$ direction the boundary conditions are fixed and the confining walls are the ones introduced above.
The number of particles in the box was kept constant at about $10^3$, ensuring comparable uncertainties at different densities.
The time integration is performed at constant temperature using the standard Nos\'e-Hoover equations~\cite{naso_84,aspirapolvere_85}, with a timestep equal to $0.001$ in reduced units.
The initial random configuration is equilibrated for about $10^8$ steps, after which a production stage of $10^9$ time steps takes place. The local properties in the $z$ direction are sampled with a granularity of $0.05\sigma$ and averaged during all the production phase. 

As shown in Section~\ref{massflow}, the local thermodynamic quantities that enter the analytical expressions for the velocity profile $v^x(z)$ and the pressure gradient are the local mass density $\rho(z)$ and the transverse component of the {\it virial} enthalpy density tensor. In inhomogeneous systems at equilibrium the enthalpy profile can not be defined unambiguously, because it depends on the pressure tensor, which can be defined in an infinite number of ways, differing by a total divergence~\cite{henderson}. This lack of uniqueness is not troubling, because all the relevant observables do not depend on the specific adopted choice.
Note that the virial transverse pressure entering our expressions for $h^v_0(z)$ is not one of the allowed definitions, and in addition it does not satisfy the hydrostatic equilibrium condition~\cite{henderson}. Nevertheless, our derivation in~\cite{pre_2022} shows that this specific expression is mathematically equivalent to the well-defined correlation function~(\ref{ss}). The transverse component of the virial pressure tensor has been evaluated using the \texttt{compute stress/atom} command in the LAMMPS package.

The velocity profile in the liquid regime, Eq.~(\ref{massflux}), is inversely proportional to the shear viscosity of the bulk fluid. This quantity can be evaluated following many different routes. Among the others, the Kubo formula relates the viscosity to the integrated dynamical correlation function of the pressure tensor at equilibrium. The correlation function has been obtained by running a periodic two-dimensional equilibrium MD simulation of a LJ fluid in the microcanonical ensemble, at the temperatures and densities of interest. The number of particles was about $2 \times 10^4$ and the simulation was run for $10^8$ steps.

\subsection{Nonequilibrium molecular dynamics simulations}
\begin{figure}
    \includegraphics{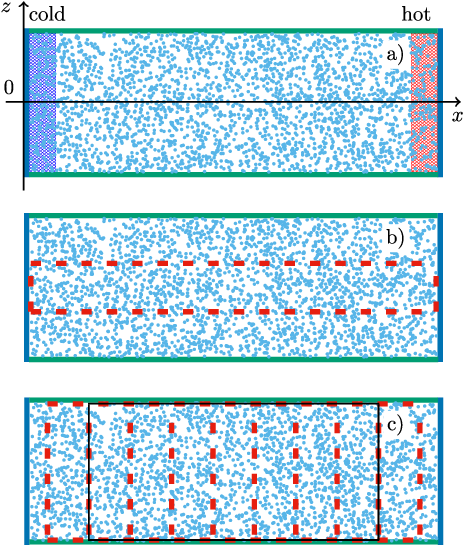}
    \caption{
    \label{fig:sys}
    Schematic of a typical system. Blue lines represent reflective walls, whereas the green ones different kind of confining surfaces. Panel a:  Blue and red shaded regions indicate the thermostated regions (cold and hot, respectively). 
    Panel b and c show the sampling scheme for the study of the nonuniform properties in the $x$ and $z$ direction respectively.}
\end{figure}

The simulation box is a two-dimensional rectangular box (see Fig.~\ref{fig:sys}a) of width $L_z=30\sigma$ and length $L_x = 180\sigma + 2 L_t$. The length of the thermostated regions $L_t$ is $30\sigma$ at low density and has been reduced to $10\sigma$ for the denser systems. Fixed boundary conditions have been applied both in the $x$ and in the $z$ direction. The number of particles in the simulation box is determined by the desired bulk density and approximately ranges from $10^2$ up to $10^4$.
Along the $x$ direction the particles are confined by purely reflective walls placed at $x=0$ and $x=L_x+2L_t$ in all the simulation runs, while
the nature of the two other confining walls at $z=-L_z/2$ and $z=L_z/2$ has been detailed before.

Each initial configuration is first equilibrated through a NVT dynamics~\cite{naso_84,aspirapolvere_85}, run for about $10^7$ steps. The time step is set equal to $0.005$ in reduced LJ units throughout all the nonequilibrium simulations.
Then, the temperature gradient $\partial_x T$ along the $x$ axis is induced by setting the temperature of the thermostated regions at two different values, $T_C$ and $T_H$, through a canonical sampling thermostat exploiting a global velocity rescaling with Hamiltonian dynamics~\cite{par_bus}. Under these conditions, a NVE time integration is performed for further $3\times10^7$ steps, after which the system reaches the steady state.
The results presented in the next Section have been obtained through an average process on the data accumulated by running the simulation for a number of steps ranging from $10^9$ to $1.2 \times 10^{11}$, depending on the bulk density (more diluted systems need larger amount of data for a better statistics).

\begin{figure}
       \input{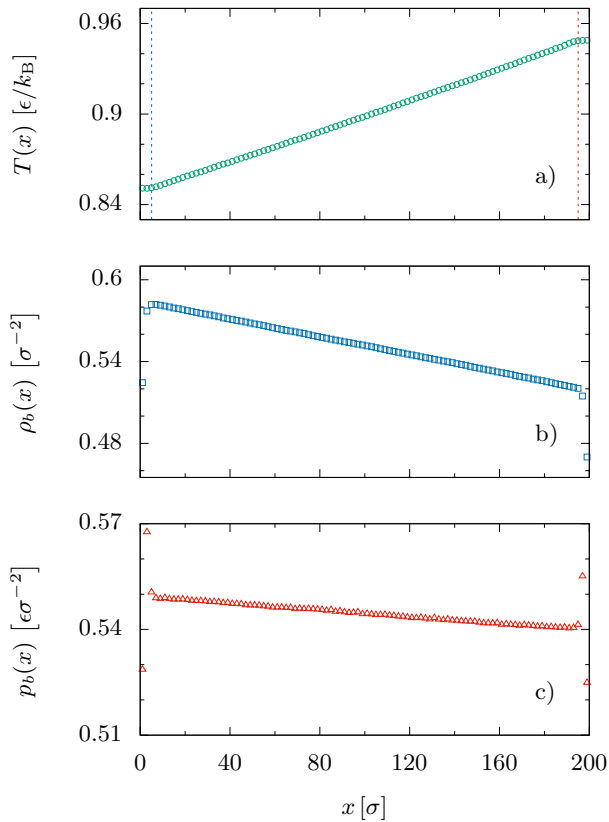}
        \caption{
        \label{fig:prof} 
        Temperature (a), density (b) and pressure (c) profile along the $x$ direction of a truncated and shifted LJ fluid confined by the potential $V_r(z)$ at $\rho_b\sigma^2=0.55$ and $T=0.9\,\epsilon/\kb$. Each point has been obtained by averaging the data at a given $x$ in the red-dashed ``bulk region" shown in Fig.~\ref{fig:sys}b.}
\end{figure}

The sampling scheme is presented in Fig.~\ref{fig:sys}. The red-dashed box in panel b is the region selected for the sampling of the properties in the $x$ direction, {\it e.g.} the bulk pressure, temperature and density profile. The granularity is $2\sigma$, the box is centered at $z=0$, the length equals $L_x$, while the height is chosen to guarantee the sampling of bulk properties. 
The bulk pressure gradient has been evaluated by extracting the slope of the linear fitting of the pressure profile (Fig.~\ref{fig:prof}c) in a region sufficiently far from the reservoir. The uncertainty on this value is the error of the fitting parameter. The pressure has been evaluated using the \texttt{compute stress/atom} command: In the red-dashed region the fluid is homogeneous, the pressure is isotropic and all the equivalent definitions in Eq.~(\ref{prob}) reduce to the virial one, implemented in this LAMMPS command. 

The $z$-dependent velocity profiles are measured in the red-dashed rectangles shown in Fig.~\ref{fig:sys}c, with a granularity in the $z$ direction of $0.05\sigma$. The velocity profile far from the reservoirs does not depend on $x$ at fixed $z$ within numerical accuracy. Therefore, smoother profiles are obtained by averaging the data in the regions inside the black rectangle in panel c.

Being interested in the linear response of the system to the perturbation induced by a temperature gradient, we have to carefully choose the temperature difference between the two reservoirs ensuring a linear behavior through the sample. Indeed, a temperature gradient $\partial_x T = 5 \times 10^{-4} \, \epsilon / (k_{\mathrm{B}} \sigma)$ gives rise to a linear profile, regardless of the bulk density, the average temperature and the external confining potential.
Figure~\ref{fig:prof}a confirms the linear behavior of the temperature profile in a system at average bulk density $0.55\,\sigma^{-2}$ confined along the $z$ direction by the external potential defined in Eq.~(\ref{eq:Vr}). In addition, within the linear response regime, also the bulk density $\rho_b$ and pressure $p_b$ should be linear functions of $x$: Our NEMD data confirm this expectation (see Fig.~\ref{fig:prof}b and~\ref{fig:prof}c).

\section{Results}
\label{sec:results}
The most relevant observables characterizing thermo-osmosis in a closed pore are the velocity profile $v^x(z)$ and the bulk pressure gradient $\partial_x p_b$, which carries the same information of the pressure drop usually measured in experiments~\cite{gaeta}. The larger the pressure gradient, the stronger is the Poiseuille backflow in the central part of the channel. In the following Sections we present the results of the pressure gradient and the velocity profile from our NEMD simulations for different confining walls and we compare them with the corresponding theoretical predictions. Here we stress a relevant point: All the expressions provided by our {\it microscopic} linear response approach, both for the pressure gradient and for the velocity profile, do not depend on free parameters, but only on the details of the interaction between particles with each other and with the external confining potential. 

\subsection{Pressure gradients}
\begin{figure}
        \input{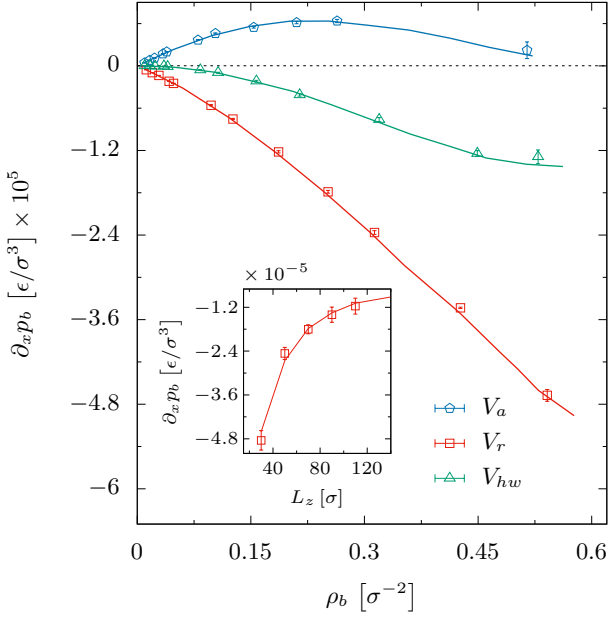}
        \caption{
        \label{fig:dpb_hw_aw_rw}
        Bulk pressure gradient of a truncated and shifted LJ fluid confined by $V_a(z)$ (blue pentagons), $V_{hw}(z)$ (green triangles),  and $V_r(z)$ (red squares) as a function of the bulk density at fixed $T=0.9\,\epsilon/\kb$ and $L_z=30\, \sigma$.
        The points have been obtained by running nonequilibrium MD simulations.
        The lines are the corresponding pressure gradients obtained from Eq.~(\ref{drop2}) without any free parameter.
        Inset: Bulk pressure gradient for the same fluid confined by $V_r(z)$ at fixed $\rho_b\sigma^2=0.54$ and $T=0.9\,\epsilon/\kb$ as a function of the channel width $L_z$.}
\end{figure}

\begin{figure}
       \input{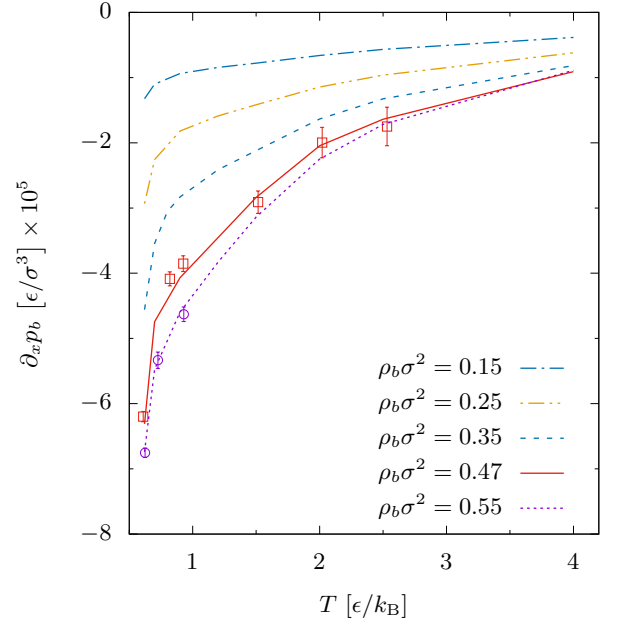}
        \caption{
        \label{fig:dpb_Tdep}
        Bulk pressure gradient of a truncated and shifted LJ fluid confined by $V_r(z)$ as a function of the average temperature at different values of the bulk density. 
        The points at $\rho_b\sigma^2=0.47$ and $\rho_b\sigma^2=0.54$ have been obtained by running nonequilibrium MD simulations.
        The lines are the corresponding pressure gradients obtained from Eq.~(\ref{drop2}).}
\end{figure}

Equation~(\ref{drop2}) provides the value of the bulk pressure in terms of the quantity $\Delta h_m$. Thus, to obtain $\partial_xp_b$ we just need to run an equilibrium MD simulation at the average density and temperature of the nonequilibrium system with a sampling of the enthalpy profile in the $z$ direction. The inspection of Figures~\ref{fig:dpb_hw_aw_rw} and~\ref{fig:dpb_Tdep} shows that our predictions are in excellent agreement with the results of the nonequilibrium MD simulation at all the density and temperature regimes of interest, for different confining potentials and channel widths.  

The simplest kind of confinement is a purely reflective wall, namely an external step potential, equal to zero in the fluid region and divergent outside. In the ideal gas limit, that is at densities where the interparticle interactions are totally negligible, this confining potential does not perturb the thermodynamic quantities, whose values coincide with the bulk ones throughout the system. Therefore Equation~(\ref{drop2}) predicts a vanishing bulk pressure gradient, that is indeed observed (see Fig.~\ref{fig:dpb_hw_aw_rw}, light-blue line and squares) at densities smaller than $0.05 \,\sigma^{-2}$. At higher densities the interaction between particles becomes more and more important and a nonzero pressure gradient in the direction opposite to the temperature gradient develops. The magnitude increases monotonically as a function of the density and seems to reach a plateau at $\rho_b\approx 0.5\,\sigma^{-2}$. 

The effect of an additional external potential has been numerically investigated by confining the fluid particles with reflective walls plus a smooth, finite-range, attractive or repulsive tail.
In this case the presence of the wall alters the enthalpy profile also in the ideal gas limit, giving rise to a non-vanishing pressure gradient, that grows linearly with the bulk density (see Fig.~\ref{fig:dpb_hw_aw_rw}, blue pentagons and red squares). The direction of $\partial_x p_b$, that coincides with the direction of the thermo-osmotic flow near the interface, is towards the cold for the attractive potential $V_a(z)$ and towards the hot for the repulsive one, and does not depend on the bulk density. The magnitude of the effect is instead strongly dependent on the bulk density: $V_r(z)$ induces a significant and monotonic increase (in absolute value) at higher densities, while the trend is non monotonic for the LJ attractive interaction.
As expected, the effect decreases as the channel width increases, as shown in the inset of Fig.~\ref{fig:dpb_hw_aw_rw}, where the results obtained from Eq.~(\ref{drop2}) are compared to the outcomes of nonequilibrium MD simulations in the case of a repulsive confinement.
 
The effect of the system's average temperature on the bulk pressure gradient has been analysed in Figure~\ref{fig:dpb_Tdep} for repulsive wall-particle interaction. Thermo-osmosis turns out to be strong at low temperatures, but its magnitude monothonically reduces at higher temperatures, independently on the value of the bulk density. This behavior is not surprising because the tails of the confining potential are scaled by the temperature in the Boltzmann weight and in the limiting case only the hard-wall contribution survives. 

\textbf{}

\begin{figure}
        \input{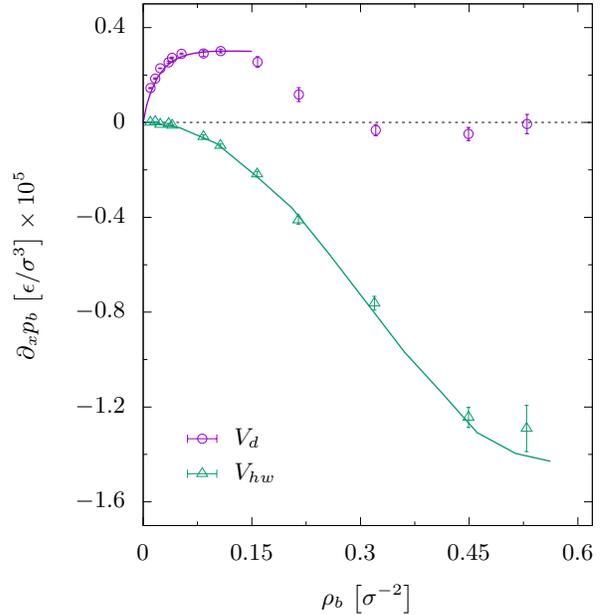}
        \caption{
        \label{fig:gpb_hw_dw}
        Bulk pressure gradient of a truncated and shifted LJ fluid confined by a reflective $V_{hw}(z)$ (green triangles - same data shown in Fig.~\ref{fig:dpb_hw_aw_rw}) and diffusive $V_d$ (violet circles) walls, as a function of the bulk density at fixed reduced temperature $0.9$.
        The points have been obtained by running nonequilibrium MD simulations.
        The lines are the corresponding pressure gradients obtained from Eq.~(\ref{drop2}) and~(\ref{drop}), with $\gamma$ evaluated in Appendix~\ref{app-diffusive}.}
\end{figure}

Figure~\ref{fig:gpb_hw_dw} shows that in the case of a diffusive surface the picture is very different. The bulk pressure gradient is positive at all the simulated densities, but it is not a monotonic function of $\rho_b$: After a linear increase at very low densities, a plateau is reached at $\rho_b = 0.05\, \sigma^{-2}$, and after a decrease it becomes negligible for densities larger than $\rho_b \approx 0.3 \,\sigma^{-2}$. 
This peculiar interface violates the momentum conservation at the surface, implying that Equation~(\ref{drop2}) is not valid. The theoretical prediction~(\ref{drop}) for the pressure gradient can still be obtained in the rarefied limit for a hard sphere fluid (see Appendix~\ref{app-diffusive}), and the agreement is remarkable until the plateau, as expected.

\subsection{Velocity profiles}
On general grounds, we expect that the physical mechanism driving thermo-osmosis, being related to interfacial effects, is not severely affected by the  boundary conditions imposed at the hot and cold reservoirs and, near the confining surfaces, the velocity profile of the fluid will not significantly depend on the channel width. 
However, we proved that a pressure gradient sets in also far from the walls, giving rise to a 
Poiseuille backflow in the bulk region, opposing to the thermo-osmotic mass transfer near the walls. Therefore we expect that the stationary velocity profile which eventually sets in will be characterized by both a {\it thermo-osmotic component} near the confining surface
and a {\it parabolic flow} in the central regions of the channel. The bulk pressure gradient will be opposite to the backflow and then will be parallel to the net mass flux in the regions close to the walls. 

The exact analytical form of the space-time dynamical correlation function ruling the thermo-osmotic velocity profile~(\ref{masscurfin}) is not available, and its direct numerical evaluation through equilibrium MD simulations is computationally very expensive.
For this reason we will compare the nonequilibrium simulations with the two approximated analytical expressions for $v_x(z)$, valid in the low~(\ref{low}) and high~(\ref{massflux}) density regimes.

\begin{figure}
        \input{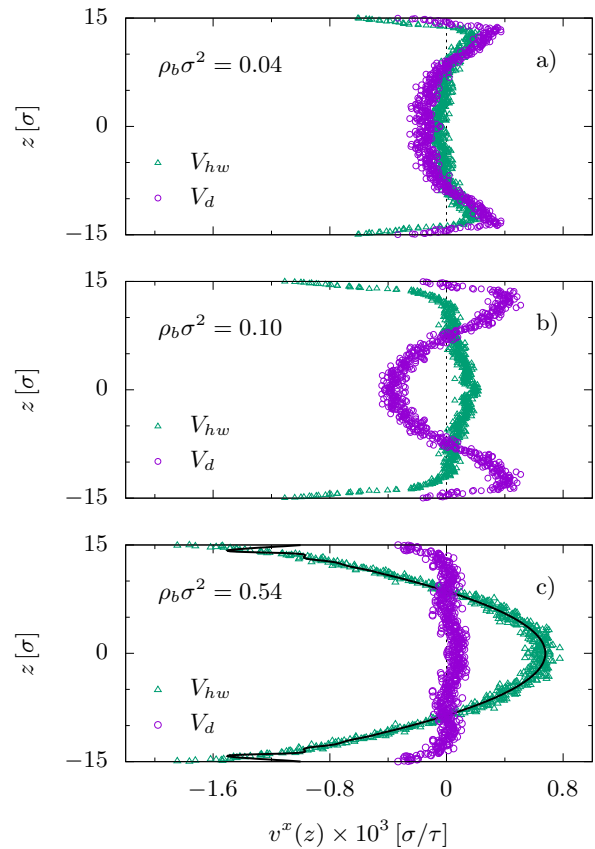}
        \caption{
        \label{fig:hdw_vel}
        Velocity profile along the nanochannel of a truncated and shifted LJ fluid confined by a purely hard wall potential (green triangles) and a diffusive potential (violet circles) at temperature $T=0.9\,\epsilon/k_{\mathrm{B}}$ at different values of the bulk density. The points have been obtained by running nonequilibrium MD simulations. The line in panel c has been obtained from Eq.~(\ref{massflux}).}
\end{figure}

We begin the overview on the results for the thermo-osmotic velocity comparing the thermo-osmotic flow arising from a simple hard wall potential and that in the case of a diffusive wall. The theoretical approach~(\ref{low}) suggests that an osmotic flow does not arise in the rarefied limit when the fluid is confined by purely reflective walls, because the enthalpy and mass densities reduce to their bulk values, and $\gamma$ equals $h_m$. Nevertheless, in panel a of Figure~\ref{fig:hdw_vel} we can notice (green triangles) that a non-vanishing thermo-osmotic flow develops near the interface also at very low density, probably because at $\rho_b = 0.04\,\sigma^{-2}$ and $\mathrm{Kn} \approx 0.6$ the intermolecular interaction is not fully negligible.
However, the absence of a backflow in the central region of the channel together with the minute bulk pressure gradient (see Fig.~\ref{fig:gpb_hw_dw}) confirm that the fluid is approaching this limit. As expected, in the case of the diffusive confinement the effect is more pronounced and at the same density an appreciable backflow towards the cold region appears. In both cases particles are driven towards the hot side, as predicted by kinetic theories~\cite{maxwell_1879}, the main difference between the two confinements being the value of the fluid velocity at contact. While in the case of the reflective wall the velocity at contact is determined only by the requirement of vanishing mass flux through the channel,
in the collision with a diffusive wall particles loose memory of the original transverse velocity, thus reducing the surface slip. As a consequence, the mass flux close to the confining surfaces is reduced for diffusive walls.
\newline
When density increases (panel b: $\rho_b = 0.1\,\sigma^{-2}$ and $\mathrm{Kn} \approx 0.2$) the differences between the two confining surfaces become apparent: At contact with the diffusive wall, the fluid velocity is still as small as before and particles are pushed towards the hot side, while the reflective wall allows the fluid to slip towards the cold side. This makes the direction of the backflow opposite for the two walls.
\newline
At even higher densities (panel c: $\rho_b = 0.54\,\sigma^{-2}$ and $\mathrm{Kn} \approx 0.04$) the system approaches the hydrodynamic regime and the thermo-osmotic effect seems to disappear in the case of the diffusive wall, whereas becomes more and more pronounced for the purely hard-wall confinement. In this limit our Equation~(\ref{massflux}) provides an excellent estimate of the velocity profile sufficiently far from the confining walls. Notice that in this expression the viscosity is a fixed parameter determined by an equilibrium MD simulation and its value is $0.781$ in reduced units. 

\begin{figure}
        \input{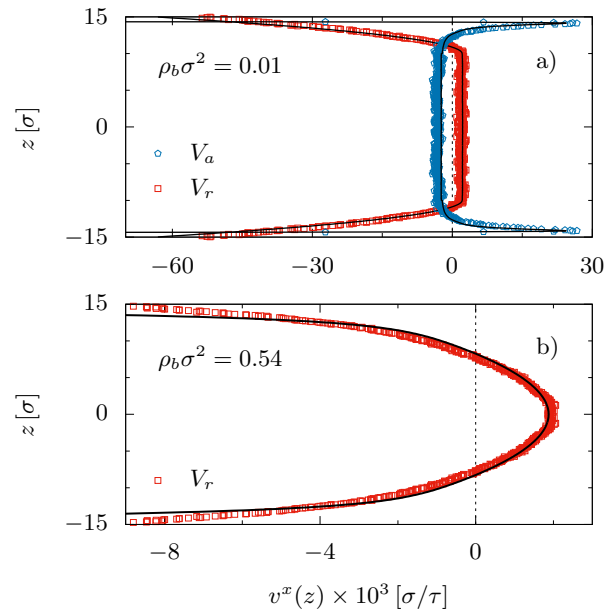}
        \caption{
        \label{fig:p_vel}
        Velocity profile of a LJ fluid confined by $V_{r}(z)$ (red squares) and $V_{a}(z)$ (blue pentagons).
        Panel a:  Rarefied limit ($\rho_b = 0.01\,\sigma^{-2}$); panel b: High-density regime ($\rho_b = 0.54\,\sigma^{-2}$). The points have been obtained by running nonequilibrium MD simulations. The lines in panel a and b have been obtained from Eq.~(\ref{low}) and (\ref{massflux}) respectively.} 
\end{figure}

The effect of the external potential on the thermo-osmotic velocity is shown in Figure~\ref{fig:p_vel}, where two representative flow profiles at low and high density are shown. Panel a refers to the low density limit $\rho_b \sigma^2 = 0.01$ and $\mathrm{Kn} \approx 2.4$ for repulsive (red squares) and attractive (blue pentagons) surfaces. The direction of the fluid is opposite in the two cases, suggesting that thermo-osmosis drives the fluid towards the cold (hot) side for repulsive (attractive) potentials. In both cases the velocity field develops within the range of the interaction and becomes flat in the central region of the channel (resembling a {\it plug} flow). The main differences between the two interfaces appear near the confining surfaces, where $V_a(z)$ (see Eq.~(\ref{eq:Va})) induces a negative velocity very close to the surface, probably due to the repulsive nature of the interaction at short range. However, this peak does not bear a significant contribution to the mass flow, because the density is very low in that small layer. The black lines superimposing the data in panel a is the result from the ideal gas approximation of our linear response approach~(\ref{low}). The enthalpy profile in~(\ref{low}) has been obtained by running an equilibrium MD simulation at the system's average density and temperature. The collision time $\tau$, appearing in the approximated expression for the velocity profile in the rarefied limit is evaluated in terms of the average thermodynamic variables~(\ref{tauave2d}) as shown in Appendix~\ref{app-collision}, where some subtle point is thoroughly discussed. Finally, we point out that the presence of a tail in the confining potential increases the thermo-osmotic effect by more than one order of magnitude with respect to the hard wall case, reported in Fig.~\ref{fig:hdw_vel}a.
\newline
In panel b we show only the velocity profile for the repulsive interaction because at high density the fluid flow for the LJ attractive potential becomes very small, with a significant noise due to poor statistics. Note that also for the repulsive wall the velocity is one order of magnitude smaller with respect to the low-density case and the profile resembles that obtained in the case of purely reflective walls (see Fig.~\ref{fig:hdw_vel}c): Far from the surfaces the backflow is parabolic in accordance with the Poiseuille law.

\begin{figure}
        \input{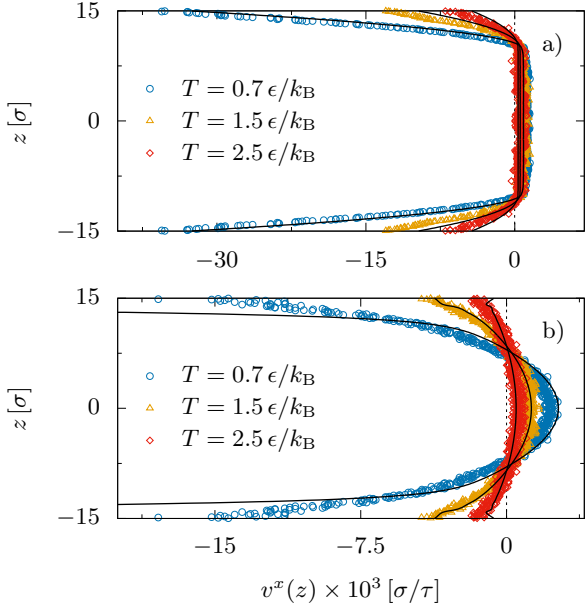}
        \caption{
        \label{fig:p_temp}
        Velocity profile of a LJ fluid confined by $V_r(z)$ at different values of the average temperature: Blue circles, $T = 0.7 \,\epsilon/\kb$; yellow triangles, $T = 1.5 \, \epsilon /\kb$; red rhombi, $T =2.5 \, \epsilon /\kb$. Panel a: Diluted system ($\rho_b = 0.03 \,\sigma^{-2} $); panel b: Dense system ($\rho_b =0.54 \,\sigma^{-2} $). The points have been obtained by running nonequilibrium MD simulations. The lines in panel a and b have been obtained from Eq.~(\ref{low}) and (\ref{massflux}) respectively. The viscosity in~(\ref{massflux}) has been evaluated through equilibrium MD simulations and equals (in reduced units) $0.734$, $0.891$ and $1.055$, from the lower to the higher temperature.}
\end{figure}

As already pointed out in the discussion of the pressure gradient, the average temperature of the system strongly influences the magnitude of the effect. In Figure~\ref{fig:p_temp} we show the velocity profile in the rarefied (panel a) and in the dense (panel b) regimes at three different average temperatures when the confining potential is $V_r(z)$. At low density the velocity has a plug form at all the temperatures and the main difference is the amount of slip near the interface, that increases at low temperatures. At high density the flow is parabolic in the central region of the channel. Again, the parameter-free theory well reproduces the simulation data. 

\section{Conclusions}
\label{conclusions}
In this paper, we conducted extensive nonequilibrium numerical simulations of thermo-osmosis for a simple fluid in a two-dimensional slab geometry. A crucial feature of our model is the choice of the confining potential: A wall-particle interaction $V(z)$ which prevents fluid-wall momentum transfer along $x$. Physically, this condition can be realized on very smooth surfaces where the $x$ component of the total momentum of the fluid is approximately conserved in the scattering event. In such an idealized system, thermo-osmosis is {\it the only mechanism} leading to momentum transfer at the wall surface, allowing for the unambiguous identification of the physical origin of the thermo-osmotic flow. 

By use of linear response theory and equilibrium statistical mechanics the {\it exact} expression for the pressure gradient which develops in a nanochannel connecting two reservoirs at different temperatures has been derived. The analytical result, expressed in terms of equilibrium properties, has been compared with nonequilibrium numerical simulations on the same model, displaying remarkable agreement. 
The velocity profile of the fluid in the same conditions is written in terms of a suitable dynamical correlation function at equilibrium. Such a correlation function can be approximately evaluated in the two limiting regimes of rarefied gas and dense fluid. In both cases the analytical velocity profile, again without free parameters, quantitatively reproduces the numerical simulation data. 

Our approach puts into a firm basis the phenomenological theories derived within nonequilibrium thermodynamics~\cite{derjaguin_sidorenkov_1941thermoosmosis}, providing the unique microscopic definition of the thermodynamic quantities of interest, like heat flux and specific enthalpy, determining the mass current in nanochannels.

The analytical expressions obtained in this study, validated by the comparison with nonequilibrium molecular dynamics simulations, pave the way to a fully microscopic interpretation of single particle thermophoresis in colloidal suspensions~\cite{pp,wurger_2010}, relating in a quantitative way this phenomenon to the character of the interactions between the solvent and the surface of the colloidal particle. As a side result of this study, we confirm that thermophoresis is driven by interfacial effects, strongly suggesting that the thermophoretic velocity of a colloidal particle in a thermal gradient can be expressed in terms of the thermo-osmotic coefficient (\ref{beta12}). As such, it turns out to be independent of the particle's radius. This observation provides a valuable contribution to the ongoing debate on this  subject~\cite{2006_braun,2007_piazza_size_soret,2008_PRL_piazza,2023_braun}.

The structure and the properties of the confining wall adopted in this study are admittedly quite simple and cannot reproduce all the details of more realistic models. However, the remarkably accurate, parameter free, comparison between theory and simulations proves in a clear-cut way that the physical mechanism underlying the thermal forces in fluids has been correctly identified.

\section*{Acknowledgements}
We gratefully thank Matteo Bessega, Roberto Piazza and Stefano Buzzaccaro for constant encouragement and illuminating discussions.

We acknowledge the CINECA award under the ISCRA initiative, for the availability of high performance computing resources and support.

We acknowledge financial support under the National Recovery and Resilience Plan (NRRP), Mission 4, Component 2, Investment 1.1, Call for tender No. 104 published on 2.2.2022 by the Italian Ministry of University and Research (MUR), funded by the European Union - NextGenerationEU - Project Title  ``Thermal Forces in confined fluids and soft solids'' - CUP J53D23001310006 - Grant Assignment Decree No. 957 adopted on 6.30.2023 by the Italian Ministry of University and Research (MUR).

\appendix
\section{Transverse virial pressure}
\label{app-virial}
Here we show that in a fluid confined in a slab, the spatial integral of the thermodynamic pressure equals the 
area of the walls times the spatial integral along the direction of confinement ($z$) of the transverse component of the virial pressure.

In a homogeneous system, the logarithm of the Grand Canonical partition function in the thermodynamic limit is expressed in terms of the pressure as: $\log \Xi = \beta PV$. In a slab of height $L_z$ and width $L_x\to\infty, L_y\to\infty$, the logarithm of the partition function will be 
proportional to $A=L_xL_y$ (in three dimensions) with a non-trivial dependence on $L_z$ (which remains finite). In this case we can define a 
``thermodynamic pressure" from 
\begin{equation}
\log \Xi = \beta P(L_z)\,A\,L_z.
	\label{ext}
\end{equation}
The expression of the Grand Canonical partition function reads:
\begin{equation}
	\Xi(A)  = \sum_N \frac{z^N}{N!} \int_{A\times \mathbb{R}} \!\! \ud\bq_1 \dots \ud\bq_N \,\ue^{-\beta \hat H(\{\bq_i\})},
 \notag
\end{equation}
where $z=\exp{(\beta \mu)}$ is the fugacity and we limit the integration over $q^x_i$ and $q^y_i$ to the domain $q^x\in(0,L_x)$, $q^y\in (0,L_y)$, whereas the integration over $q^z_i$ is extended to the whole real domain. The finite height $L_z$ of the system is encoded in the external potential $V(z)$ included in $\hat{H}$. Now we evaluate the change in the partition function when $L_x\to L_x\,(1+\epsilon)$: 
\begin{align}
	&\Xi(A_\epsilon) \! = \!\sum_N \frac{z^N}{N!}\!\!\int_{A_\epsilon\times \mathbb{R}} \!\ud\bq_1 \dots \ud\bq_N \,\ue^{-\beta \hat H\left(\left\{q^x_i,q^y_i,q^z_i\right\}\right)}  \nonumber \\
	&= \, \sum_N \frac{z^N(1+\epsilon)^N}{N!} \!\!\!\,\,\int_{A\times \mathbb{R}} \!\ud\bq_1 \dots \ud\bq_N \,\ue^{-\beta \hat H\left(\left\{q^x_i(1+\epsilon),q^y_i,q^z_i\right\}\right)},\nonumber
\end{align}
where $A_\epsilon = A\,(1+\epsilon)$ and a rescaling $q^x_i\to q^x_i(1+\epsilon)$ has been performed in the second line. Expanding to first order we get:
\begin{eqnarray}
	\Xi(A_\epsilon) &=& \Xi(A)+\epsilon\,\Bigg [ 
	\sum_N N\,\frac{z^N}{N!} \int_{A\times \mathbb{R}}\!\!\! \ud\bq_1 \dots \ud\bq_N \,\ue^{-\beta \hat H(\{\bq_i\})} \nonumber \\ 
	&& -\beta \, 
	\sum_N \!\frac{z^N}{N!}\! \int_{A\times \mathbb{R}}\!\!\! \ud\bq_1 \dots \ud \bq_N \,\ue^{-\beta H(\bq_i)}\, \!\! \sum_i q^x_i\,\frac{\partial \hat H}{\partial q^x_i}\Bigg ] \nonumber \\
	&=& \Xi(A)\,\left[ 1+ \epsilon \,\left( \bla \hat N\bra -\beta \left\langle \! \sum_i q^x_i\,\frac{\partial \hat H}{\partial q^x_i}\right\rangle \right) \right].
	\nonumber 
\end{eqnarray}
Therefore 
\begin{equation}
	L_x\,\frac {\partial \log\Xi}{\partial L_x} = \bla \hat N\bra -\beta \left \langle  \sum_i q^x_i\,\frac{\partial \hat H}{\partial q^x_i}\right \rangle.
 \notag
\end{equation}
From the relation~(\ref{ext}) we then obtain
\begin{equation}
	P(L_z) = V^{-1} \, \left ( \bla \hat N\bra \,\kb T -\bigg\langle  \sum_i q^x_i\,\frac{\partial \hat H}{\partial q^x_i} \bigg\rangle \right ),
 \notag
\end{equation}
where $V=AL_z$, which coincides (after some straightforward algebra) with the spatial average of the transverse component of the (local) {\it virial} pressure $p^v_0(z)$:
\begin{eqnarray}
	p^v_0(z) &=& n_0(z)\,\kb T - \notag \\
	&& \frac{1}{2} \int \ud\br^\prime \, n_2(\br,\br^\prime)\,
\frac{\ud v(s)}{\ud s} \,\frac{(x-x^\prime)^2}{s}\bigg\vert_{s=|\br-\br^\prime|} .
\nonumber
\end{eqnarray}
Then $ P(L_z) \,L_z = \int \ud z\, p^v_0(z) $ and the thermodynamic enthalpy $H(L_z)$ of the confined system is 
\begin{equation}
	H(L_z)=U(L_z)+P(L_z)V=A\int \ud z\, h^v_0(z) \, ,
 \label{eq:H_lz}
\end{equation}
where the transverse component of the (local) {\it virial} enthalpy reads
\begin{align}
 &h_0^v(z)=\left [ \frac{5}{2}\kb T +V(z)\right ] n_0(z) \label{eq:virial_enthalpy}  \\
& \quad   +\frac{1}{2} \int \ud\br^\prime \, n_2(\br,\br^\prime)
\left [ v(s) -  \frac{\ud v(s)}{\ud s} \,\frac{(x-x^\prime)^2}{s}\right ]_{s=|\br-\br^\prime|}. \nonumber 
\end{align}

\section{Collision time}
\label{app-collision}
Here we want to evaluate the average collision time between two hard particles within elementary kinetic theory. 

Let us consider two hard spheres of diameter $\sigma$ moving in the space. At time $t$ the centers of the two spheres are located in $\br$ and $\br_1$ with
$|\br_1-\br| > \sigma$. If the two particles have velocieties $\bv$ and $\bv_1$ respectively, after a time interval $\Delta t$ the two spheres move
in $\br+\bv\,\Delta t$ and $\br_1+\bv_1\,\Delta t$ and will suffer a collision if $|\br_1+\bv_1\,\Delta t - \br-\bv\,\Delta t| < \sigma$.
Expanding for small $\Delta t$ the condition reads:
$|\br_1-\br| + \Delta t\, (\br_1-\br)\cdot (\bv_1-\bv)/|\br_1-\br| < \sigma$.
Then the collision probability in the time interval $\Delta t$ equals the probability that the initial position of the two particles  belong
to the interval
\begin{equation}
\sigma < |\br_1-\br| < \sigma - \Delta t\, |\bv_1-\bv|\,\cos\theta,
        \label{anulus}
\end{equation}
where $\theta$ is the angle between $\br_1-\br$ and $\bv_1-\bv$. This means that $\cos\theta$ must be negative.
Fixing the position $\br$ and the two velocities $\bv$ and $\bv_1$,
the probability to find a particle in $\br_1$ satisfying this condition equals the number density $n$ times
the volume defined by Eq.~(\ref{anulus}):
\begin{eqnarray}
        \Delta V &=& \int_0^{2\pi} \ud \varphi \int_{\frac{\pi}{2}}^{\pi} \ud\theta\,\sin\theta \int_\sigma^{\sigma - \Delta t\, |\bv_1-\bv|\,\cos\theta} \ud R\, R^2 \nonumber \\
        &=&\pi\,\sigma^2 \, |\bv_1-\bv|\,\Delta t.
        \notag
\end{eqnarray}
This probability is usually written as
$\frac{\Delta t}{\tau}$, leading to the definition of collision rate:
\begin{equation}
        \frac{1}{\tau} = \pi\,n \sigma^2 \,|\bv_1-\bv|.
        \notag
\end{equation}
From this expression we can evaluate the averages with respect to the velocity distribution of the two particles, either averaging the collision frequency or the collision time. The velocity distribution is given by the equilibrium Maxwell-Boltzmann form:
\begin{equation}
        f(v) = \left (\frac{\beta m}{2\pi}\right )^{\frac{3}{2}} \,\ue^{-\frac{1}{2}\beta m v^2} \, .
        \notag
\end{equation}
In order to average the collision frequency we have to evaluate
\begin{equation}
        \bla \tau^{-1} \bra = \int \ud\bv \int \ud\bv_1 \, f(v)\,f(v_1) \, \pi n \sigma^2 \,|\bv_1-\bv| .
        \notag
\end{equation}
By first introducing the center of mass velocity $\bC$ and the relative velocity $\bV$ via $\bv = \bC +\frac{1}{2}\bV$ and
$\bv_1 = \bC -\frac{1}{2}\bV$, the integral can be easily performed with the result:
\begin{equation}
\bla \tau^{-1} \bra = 4n\sigma^2\sqrt{\frac{\pi}{\beta m}}, \notag
\end{equation}
which coincides with the expression reported in textbooks~\cite{balescu}.
Instead, averaging the collision time, the same steps lead to 
\begin{equation}
\langle \tau\rangle = \frac{1}{\pi n\sigma^2}\,\sqrt{\frac{\beta m}{\pi}} = \frac{4}{\pi} \,\frac{1}{\bla \tau^{-1}\bra}.
\label{tauave3d}
\end{equation}
In our expression~(\ref{low}) the mass flux is proportional to 
$\tau$ and then it is natural to use the latter estimate 
(\ref{tauave3d}). 

Molecular dynamics simulations have been performed in two dimensions, therefore here we report also the analogous expressions appropriate for hard disks:
\begin{eqnarray}
\bla \tau^{-1} \bra &=& 4n\sigma^2\sqrt{\frac{\pi}{\beta m}} \, ; \notag \\
\langle \tau\rangle &=& \frac{\sqrt{\pi\beta m}}{4 n \sigma} =\frac{\pi}{2} \,\frac{1}{\bla \tau^{-1}\bra} \, .
\label{tauave2d}
\end{eqnarray}

\section{Diffusive wall}
\label{app-diffusive}
The diffusive wall violates momentum conservation and then the continuity equation~(\ref{cont}) is no longer valid {\it on} the confining surface. 
In a rarefied gas,
elementary kinetic theory allows to explicity evaluate the time derivative of the quantity $F(t) = \int \ud z \,F(z,t)$ (see Eq.~(\ref{f})),
which is now non-vanishing due to the loss of 
momentum during particle-wall scattering. In the rarefied limit, only the kinetic terms contribute to the definition of the heat current and
$F(t)$ can be explicitly written as: 
\begin{equation}
	F(t)  = \frac{1}{A} \left \langle \sum_i p_i^x(t) \,\sum_j  \frac{p^x_j}{m}\left [ \frac{p_j^2}{2m}-\gamma m\right ] \right \rangle_0,
	\label{gxx3}
\end{equation}
where $A$ is the area of the $(x,y)$ section. The change in $F(t)$ in the interval $(t,t+\Delta t)$ is due to the change in the total momentum 
(the first factor in Eq.~(\ref{gxx3})) in $\Delta t$ and comes from the particle-wall scattering alone. Each particle hitting the wall in $\Delta t$
and having a velocity $v^z$ in the orthogonal direction belongs to the volume $\Delta V = A\,|v^z| \,\Delta t$ where the sign of $v^z$ must be 
negative at the bottom surface and positive at the upper surface. On average, each of these particles loses its momentum $p^x$ during the scattering. 
Moreover, the only contribution to this correlation function comes from those particles which have not experienced intermolecular collisions and then move ballistically up to time $t$, therefore
\begin{equation}
	\frac{\Delta F(t)}{\Delta t}  = -\frac{\rho_0}{m}\,\left \langle \left|v^z\right|\,p^x\,
\frac{p^x}{m}\left [ \frac{p^2}{2m}-\gamma m\right ] \right \rangle_0.
\notag
\end{equation}
The average is limited to those particles reaching either surface at time $t$. 
If $\tau$ is the collision time, the survival probability at time $t$ is $\ue^{-t/\tau}$ and, performing the average, we obtain 
\begin{eqnarray}
	\frac{\ud F(t)}{\ud t} &=& -\frac{2\,\rho_0\, \ue^{-t/\tau}}{\sqrt{2\pi}} 
	\left (\frac{\kb T}{m}\right)^{\frac{3}{2}}\,  \kb T\nonumber \\
	&& \left[ (3 -\beta\gamma\,m)\,\left(1-\ue^{-\xi^2}\right) -\xi^2\,\ue^{-\xi^2}\right], \quad
	\label{derg}
\end{eqnarray}
where $\xi^2 = {\beta m\,L_z^2}/({2\,t^2})$ and $L_z$ is the width of the channel. We stress that, when the $x$ component of the momentum is not conserved in a wall-particle scattering event, the quantity $F(t)$ does depend on time, contrary to Eq.~(\ref{f00}), valid for passive walls.

\begin{figure}
    \input{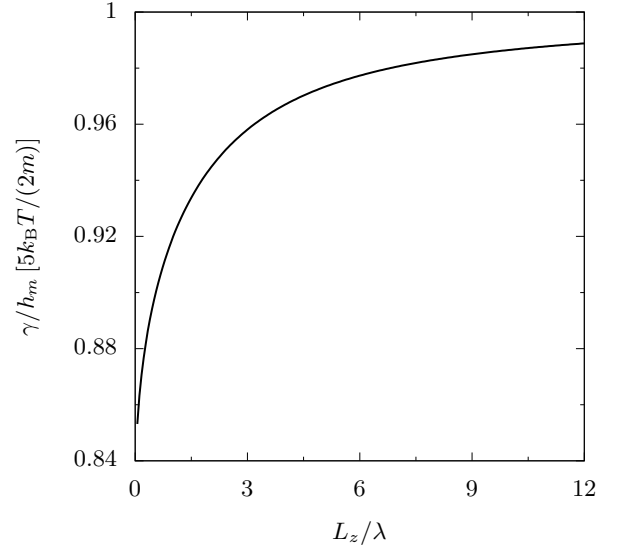}
    \caption{
    Solution of Eq.~(\ref{eqgamma}) for $\gamma$, in units of $h_m=\frac{5}{2} \frac{\kb T}{m}$, as function of	the channel width $L_z$, in units of the mean free path $\lambda$.
    \label{fig-gamma}}
\end{figure}

Integrating in time Eq.~(\ref{derg}) with the condition that $F(t)$ must vanish as $t\to\infty$, we get:
$F(t) = -\int_t^\infty \ud t^\prime \, \Gamma(t^\prime)$ where $\Gamma(t)$ is the right hand side of Eq.~(\ref{derg}). 
However,  $F(0)$ is just a static correlation function which can be exactly evaluated.
Therefore we can write the equation determining the value of the parameter $\gamma$: 
\begin{eqnarray}
	F(0) &=& L_z \,\rho_0 \,(\kb T)^2\, \left [ \frac{5}{2m} - \beta\gamma\right ] \nonumber \\
&=&  -\int_0^\infty \ud t^\prime \, \Gamma(t^\prime) \, .
	\label{eqgamma}
\end{eqnarray}
This equation simplifies for a wide channel, much larger than the mean free path $\lambda$. 
In this case, $\xi$ can be pushed to infinity in Eq.~(\ref{derg}) 
and Eq.~(\ref{eqgamma}) simply gives $\gamma = 5\kb T / (2m)=h_m$ plus corrections $O\left (\lambda/L_z\right )$. 
A plot of the predicted value of $\gamma$ as a function of the channel width is depicted in Fig.~\ref{fig-gamma}.
As shown in Eq.~(\ref{drop}), his quantity is directly related to the pressure difference $\Delta p$ between the hot and the cold side of the channel. 
Therefore, Fig.~\ref{fig-gamma} allows to directly evaluate the pressure gradient by use of Eq.~(\ref{drop}).

In order to impose the  vanishing of the integrated mass flux, we cannot set the velocity shift $u_0^x$ to zero as in the case of passive walls
in Eq.~(\ref{masscurfin}). The appropriate value of $u_0^x$ is determined by the equation 
$\rho_0 L_z\, u_0^x + \int_0^\infty \ud t \, F(t)\partial_x\beta = 0$, expressing mass conservation. 
Integrating by parts and using Eq.~(\ref{derg}) we find
\begin{eqnarray}
	u_0^x &=& -\frac{(\kb T)^2}{\sqrt{\pi}\,m}\, \partial_x\beta\,\int_0^\infty \ud t \,\ue^{-t/\tau}\nonumber \\
	&&\xi^{-1}\,\left [ (3 -\beta\gamma\,m)\,\left(1-\ue^{-\xi^2} \right) -\xi^2\,\ue^{-\xi^2}\right ].
\label{uzero}
\end{eqnarray}
The shift vanishes as $L_z^{-1}$ for large channels. 

Having determined $\gamma$ from Eq.~(\ref{eqgamma}) and $u_0^x$ from Eq.~(\ref{uzero}), we can now evaluate the velocity profile.
In order to be consistent with the previous analysis, we will adopt elementary kinetic theory to calculate the time derivative of the mass current-heat current dynamical correlation: 
\begin{equation}
	\frac{\partial}{\partial t} \!\int\! \ud\br^\prime \bla \hat j^x(\br,t) \hat J^x_Q(\br^\prime) \bra_0
	= - \partial_z \! \int \!\ud\br^\prime\bla \hat J^{xz}(\br,t) \hat J^x_Q(\br^\prime) \bra_0. \nonumber 
\end{equation}
Here we used the continuity equation which remains valid inside the channel, that is for $ -L_z/2 < z < {L_z}/{2}$, 
except precisely at the confining walls $z=\pm \, L_z/2$. 
In a rarefied gas, the correlation function at right hand side can be explicitly written as
\begin{eqnarray}
	&& \int \ud\br^\prime \,\bla \hat J^{xz}(\br,t)\, \hat J^x_Q(\br^\prime) \bra_0 = \frac{N(t)}{2 m^3 A}\nonumber \\
&& \qquad\bla p^x(t) p^z(t) \,\delta\big(z-z(t)\big)\,p^x\big( p^2 - 2m^2 \gamma \big)\bra_0 \,,\quad
	\label{disp}
\end{eqnarray}
where $z(t)=z_0 + p^zt/m $ is the coordinate of the particle at time $t$, while $z_0$ is the position at $t=0$ 
and $N(t)=N\,\ue^{-t/\tau}$ is the number of particles which did not experience intermolecular collisions up to time $t$. 
In the case of a purely reflective hard wall, the average vanishes because $i)$ the particle is 
supposed to move ballistically and cannot have suffered collisions from other particles in the gas in the interval $[0,t]$ and $ii)$ $p^x$ and $p^2$ are conserved during the reflection at the wall. Then the average can be performed over the $z$ component of the position and of the momentum of the particle at time $t$, but Eq.~(\ref{disp}) is odd in $p^z(t)$: The average vanishes and then no particle flow sets in.
Instead, for a diffusive wall, only particles which do not undergo scattering against the wall in the interval $[0,t]$ move in a ballistic way
and conserve the $x$ component of their momentum. The direct evaluation of this average gives: 
\begin{eqnarray}
        && \int \ud\br^\prime \,\bla \hat J^{xz}(\br,t)\, \hat J^x_Q(\br^\prime) \bra_0 = 
	-\frac{\rho_0}{\sqrt{2\pi}}\,(\kb T)^{\frac{5}{2}}m^{-\frac{3}{2}}\, \ue^{-t/\tau}\nonumber \\
	&& \qquad \left [ (3-\beta\gamma m) \left(\ue^{-\zeta_+^2}-\ue^{-\zeta_-^2}\right) + 
	\zeta_+^2\ue^{-\zeta_+^2}-\zeta_-^2\ue^{-\zeta_-^2}\right ], \notag
\end{eqnarray}
with $\zeta_\pm^2 = {\beta m(L_z\pm 2z)^2}/ \left(8t^2\right)$. It is straightforward to check that this expression is fully consistent with the evaluation of the time derivative of $F(t)$ performed in Eq.~(\ref{derg}). Taking the $z$ derivative and integrating in time, the velocity profile can be obtained from Eq.~(\ref{masscurfin}) and is given in Eq.~(\ref{veldif}).

\end{document}